\documentclass[12pt]{article}

\pdfoutput=1

\usepackage{amsmath,amssymb,amsfonts,amscd,mathrsfs}
\usepackage{xcolor}
\definecolor{darkblue}{rgb}{0.1,0.1,.7}
\usepackage[colorlinks, linkcolor=darkblue, citecolor=darkblue, urlcolor=darkblue, linktocpage]{hyperref} 
\usepackage{geometry}
\usepackage{tablefootnote,colortbl}

\usepackage{braket}

\usepackage{cancel}

\usepackage{arydshln}
\usepackage{booktabs,multirow}

\usepackage{tikz-feynman}
 \tikzfeynmanset{compat=1.0.0}
 
 \usepackage[square, comma, compress,numbers]{natbib}
\definecolor{myorange}{RGB}{199,146,32}

\definecolor{Gray1}{gray}{0.97}
\definecolor{Gray2}{gray}{0.9}
\definecolor{LightCyan}{rgb}{0.88,1,1}
\definecolor{blu}{rgb}{0,0,1}

\newcommand{\diagTWO}{
  \begin{minipage}[h]{0.12\linewidth}\begin{tikzpicture}
  [
roundnode/.style={circle, draw=black!60, fill=black!6, very thick, %minimum size=7mm,
  inner sep=1pt,
  text width=3mm},
]
\begin{feynman}[small]
\vertex (X) at (0,0);
\vertex (Xaux) at (0,-.1);
\vertex (x1l) at (-.7,.6);
\vertex (x1lp) at (-.7-.15,.6-.02){\footnotesize$^{1}$};
\vertex (x2l) at (-.7,.35);
\vertex (x2lp) at (-.7-.15,.35-.07){\footnotesize$^{2}$};
\vertex (xnl) at (-.7,-.6);
\vertex (xnlp) at (-.7-.15,-.6-.05){\footnotesize$^{n}$};
\vertex (x1r) at (+.7,.4);
\vertex (x2r) at (+.7,.15);
\vertex (xnr) at (+.7,-.4);
\vertex (c1) at (.8,.45);
\vertex (c2) at (.8,-.6);
\vertex (Y) at (0+1.6,0);
\vertex (Yaux) at (0+1.6,-.1);
\vertex (y1l) at (-.3+1.2,.4);
\vertex (y2l) at (-.3+1.2,.15);
\vertex (ynl) at (-.3+1.2,-.4);
\vertex (D1) at (-.67,.05){$\vdots$};
\vertex (D2) at (.55,.0){$\vdots$};
\vertex (D3) at (1.05,.0){$\vdots$};
   \diagram*{
   (x1l) -- [thick, quarter left, looseness=.8] (X) --[ thick, quarter left, looseness=.8] (x1r) ,
      (x2l) -- [thick, quarter left,looseness=.8] (Xaux) --[ thick, quarter left, looseness=.8] (x2r) ,
   (xnl) -- [thick, quarter right, looseness=.8] (X) --[ thick, quarter right, looseness=.8] (xnr) ,
   (y1l) -- [thick, quarter left, looseness=.8] (Y)   ,
      (y2l) -- [thick, quarter left, looseness=.8] (Yaux)   ,
   (ynl) -- [thick, quarter right, looseness=.8] (Y)   ,
   (c1) -- [very thick, scalar, red] (c2),
};
\node[roundnode] (Y1) at (0+1.6,0){\tiny{$O_i$}};
\node[roundnode] (X1) at (0,0){\tiny{$iT$}};
  \end{feynman}
\end{tikzpicture}
  \end{minipage} 
  }

\newcommand{\diagTHREE}{
  \begin{minipage}[h]{0.12\linewidth}\begin{tikzpicture}
  [
roundnode/.style={circle, draw=black!60, fill=black!6, very thick, %minimum size=7mm,
  inner sep=1pt,
  text width=3mm},
]
\begin{feynman}[small]
\vertex (X) at (0,0);
\vertex (Xaux) at (0,-.1);
\vertex (x1l) at (-.7,.4);
\vertex (x1lp) at (-.7-.15,.4-.02){\footnotesize$^{1}$};
\vertex (x2l) at (-.7,.2);
\vertex (x2lp) at (-.7-.15,.2-.07){\footnotesize$^{2}$};
\vertex (xnl) at (-.7,-.4);
\vertex (xnlp) at (-.7-.15,-.4-.05){\footnotesize$^{k}$};
\vertex (x1r) at (+.5,.3);
\vertex (x2r) at (+.5,-.3);
\vertex (c1) at (.6,.35);
\vertex (c2) at (.6,-.4);
\vertex (Y) at (0+1.2,0);
\vertex (Yaux) at (1.2,-.1);
\vertex (y1l) at (-.3+1.,.3);
\vertex (y2l) at (-.3+1.,-.3);
\vertex (D1) at (-.67,.01){$\vdots$};
\vertex (D2) at (1.87,.01){$\vdots$};
\vertex (X1l) at (1.9,.4);
\vertex (X1lp) at (1.9+.34,.4-.02){\footnotesize$^{k+1}$};
\vertex (X2l) at (1.9,.2);
\vertex (X2lp) at (1.9+.34,.2-.07){\footnotesize$^{k+2}$};
\vertex (Xnl) at (1.9,-.4);
\vertex (Xnlp) at (1.9+.19,-.4-.05){\footnotesize$^{n}$};
   \diagram*{
   (x1l) -- [thick, quarter left, looseness=.8] (X) --[ thick, quarter left, looseness=.8] (x1r) ,
      (x2l) -- [thick, quarter left,looseness=.8] (Xaux) ,
         (xnl) -- [thick, quarter right, looseness=.8] (X) --[ thick, quarter right, looseness=.8] (x2r) ,
   (y1l) -- [thick, quarter left, looseness=.8] (Y) -- [thick, quarter left, looseness=.8] (X1l)   ,
   (y2l) -- [thick, quarter right, looseness=.8] (Y),
   (Yaux)   -- [thick, quarter left, looseness=.8] (X2l)  ,
    (Y)   -- [thick, quarter right, looseness=.8] (Xnl)  ,
   (c1) -- [very thick, scalar, red] (c2),
};
\node[roundnode] (Y1) at (0+1.2,0){\tiny{$O_i$}};
\node[roundnode] (X1) at (0,0){\tiny{$iT$}};
  \end{feynman}
\end{tikzpicture}
  \end{minipage} 
  }

\newcommand{\dQuatre}{
  \begin{minipage}[h]{0.12\linewidth}\begin{tikzpicture}
  [
roundnode/.style={circle, draw=black!60, fill=black!6, very thick, %minimum size=7mm,
  inner sep=1pt,
  text width=3mm},
]
\begin{feynman}[small]
\vertex (X) at (0,0);
\vertex (Xaux) at (0,-.1);
\vertex (x1l) at (-.5,.3);
\vertex (x1lp) at (-.5-.1,.3-.02){\footnotesize$^{1}$};
\vertex (xnl) at (-.5,-.3);
\vertex (xnlp) at (-.5-.12,-.3-.02){\footnotesize$^{2}$};
\vertex (x1r) at (+.5,.3);
\vertex (x2r) at (+.5,-.3);
\vertex (c1) at (.6,.35);
\vertex (c2) at (.6,-.4);
\vertex (Y) at (0+1.2,0);
\vertex (Yaux) at (1.2,-.1);
\vertex (y1l) at (-.3+1.,.3);
\vertex (yp1l) at (-.3+1.-.07,.3+.18){\footnotesize$^{1^\prime}$};
\vertex (y2l) at (-.3+1.,-.3);
\vertex (yp2l) at (-.3+1.-.07,-.3-.30){\footnotesize$^{2^\prime}$};
\vertex (D2) at (1.87,.01){$\vdots$};
\vertex (X1l) at (1.9,.4);
\vertex (X1lp) at (1.9+.19,.4-.02){\footnotesize$^{3}$};
\vertex (X2l) at (1.9,.2);
\vertex (X2lp) at (1.9+.19,.2-.07){\footnotesize$^{4}$};
\vertex (Xnl) at (1.9,-.4);
\vertex (Xnlp) at (1.9+.19,-.4-.05){\footnotesize$^{n}$};
   \diagram*{
   (x1l) -- [thick, quarter left, looseness=.8] (X) --[ thick, quarter left, looseness=.8] (x1r) ,
         (xnl) -- [thick, quarter right, looseness=.8] (X) --[ thick, quarter right, looseness=.8] (x2r) ,
   (y1l) -- [thick, quarter left, looseness=.8] (Y) -- [thick, quarter left, looseness=.8] (X1l)   ,
   (y2l) -- [thick, quarter right, looseness=.8] (Y),
   (Yaux)   -- [thick, quarter left, looseness=.8] (X2l)  ,
    (Y)   -- [thick, quarter right, looseness=.8] (Xnl)  ,
   (c1) -- [very thick, scalar, red] (c2),
};
\node[roundnode] (Y1) at (0+1.2,0){\tiny{$O_i$}};
\node[roundnode] (X1) at (0,0){\tiny{$iT$}};
  \end{feynman}
\end{tikzpicture}
  \end{minipage} 
  }

\newcommand{\LambUn}[6]{
  \begin{minipage}[h]{0.4\linewidth}\begin{tikzpicture}
  [
roundnode/.style={circle, draw=black!60, fill=black!6, very thick, %minimum size=7mm,
  inner sep=1pt,
  text width=3mm},
]
\begin{feynman}[small]
\node[dot] (X) at (0,0);
\vertex (x1l) at (-.5,.3);
\vertex (x1lp) at (-.5-.15,.3+.02){\footnotesize$^{#1}$};
\vertex (xnl) at (-.5,-.3);
\vertex (xnlp) at (-.5-.17,-.3-.02){\footnotesize$^{#2}$};
\vertex (x1r) at (+.5,.3);
\vertex (x2r) at (+.5,-.3);
\vertex (c1) at (.6,.35);
\vertex (c2) at (.6,-.4);
\node [crossed dot] (Y) at (1.2,0);
\vertex (Yaux) at (1.2,-.1);
\vertex (y1l) at (-.3+1.,.3);
\vertex (yp1l) at (-.3+1.-.07,.3+.25){\footnotesize$^{#5}$};
\vertex (y2l) at (-.3+1.,-.3);
\vertex (yp2l) at (-.3+1.-.07,-.3-.31){\footnotesize$^{#6}$};
\vertex (X1l) at (1.7,.3);
\vertex (X1lp) at (1.7+.19,.3-.02){\footnotesize$^{#3}$};
\vertex (Xnl) at (1.7,-.3);
\vertex (Xnlp) at (1.7+.19,-.3-.05){\footnotesize$^{#4}$};
   \diagram*{
   (x1r) -- [thick, quarter right,  scalar ,looseness=.8] (X) --[ thick, quarter left, looseness=.8, scalar]  (xnl),
         (x2r) -- [thick, quarter left, looseness=.8, scalar] (X) --[ thick, quarter right, looseness=.8, scalar] (x1l)  ,
  (Xnl)  -- [thick, quarter left, looseness=.8, scalar] (Y) -- [thick, quarter right, looseness=.8 , scalar]   (y1l)  ,
  (X1l)   -- [thick, quarter right, looseness=.8, scalar] (Y) -- [thick, quarter left, looseness=.8, scalar]    (y2l),
   (c1) -- [very thick, scalar, red] (c2),
};
\node[roundnode] (X1) at (0,0);%{\tiny{$iT$}};
  \end{feynman}
\end{tikzpicture}
  \end{minipage} 
  }

  \newcommand{\LambDos}[6]{
  \begin{minipage}[h]{0.4\linewidth}\begin{tikzpicture}
  [
roundnode/.style={circle, draw=black!60, fill=black!6, very thick, %minimum size=7mm,
  inner sep=1pt,
  text width=3mm},
]
\begin{feynman}[small]
\node[dot] (X) at (0,0);
\vertex (x1l) at (-.5,.3);
\vertex (x1lp) at (-.5-.15,.3+.02){\footnotesize$^{#1}$};
\vertex (xnl) at (-.5,-.3);
\vertex (xnlp) at (-.5-.17,-.3-.02){\footnotesize$^{#2}$};
\vertex (x1r) at (+.5,.3);
\vertex (x2r) at (+.5,-.3);
\vertex (c1) at (.6,.35);
\vertex (c2) at (.6,-.4);
\node [crossed dot] (Y) at (1.2,0);
\vertex (Yaux) at (1.2,-.1);
\vertex (y1l) at (-.3+1.,.3);
\vertex (yp1l) at (-.3+1.-.07,.3+.25){\footnotesize$^{#5}$};
\vertex (y2l) at (-.3+1.,-.3);
\vertex (yp2l) at (-.3+1.-.07,-.3-.31){\footnotesize$^{#6}$};
\vertex (X1l) at (1.7,.3);
\vertex (X1lp) at (1.7+.21,.3-.02){\footnotesize$^{#3}$};
\vertex (Xnl) at (1.7,-.3);
\vertex (Xnlp) at (1.7+.21,-.3-.05){\footnotesize$^{#4}$};
   \diagram*{
   (x1r) -- [thick, quarter right,  scalar ,looseness=.8] (X) --[ thick, quarter left, looseness=.8, scalar]  (xnl),
         (x2r) -- [thick, quarter left, looseness=.8, scalar] (X) --[ thick, quarter right, looseness=.8, scalar] (x1l)  ,
  (Xnl)  -- [thick, quarter left, looseness=.8,   scalar] (Y) -- [thick, quarter right, looseness=.8 , scalar]   (y1l)  ,
  (X1l)   -- [thick, quarter right, looseness=.8, scalar] (Y) -- [thick, quarter left, looseness=.8, scalar]    (y2l),
   (c1) -- [very thick, scalar, red] (c2),
};
\node[roundnode] (X1) at (0,0);%{\tiny{$iT$}};
  \end{feynman}
\end{tikzpicture}
  \end{minipage} 
  }

  \newcommand{\LambDosAL}[6]{
  \begin{minipage}[h]{0.4\linewidth}\begin{tikzpicture}
  [
roundnode/.style={circle, draw=black!60, fill=black!6, very thick, %minimum size=7mm,
  inner sep=1pt,
  text width=3mm},
]
\begin{feynman}[small]
\node[dot] (X) at (0,0);
\vertex (x1l) at (-.5,.3);
\vertex (x1lp) at (-.5-.15,.3+.02){\footnotesize$^{#1}$};
\vertex (xnl) at (-.5,-.3);
\vertex (xnlp) at (-.5-.17,-.3-.02){\footnotesize$^{#2}$};
\vertex (x1r) at (+.5,.3);
\vertex (x2r) at (+.5,-.3);
\vertex (c1) at (.6,.35);
\vertex (c2) at (.6,-.4);
\node [crossed dot] (Y) at (1.2,0);
\vertex (Yaux) at (1.2,-.1);
\vertex (y1l) at (-.3+1.,.3);
\vertex (yp1l) at (-.3+1.-.07,.3+.25){\footnotesize$^{#5}$};
\vertex (y2l) at (-.3+1.,-.3);
\vertex (yp2l) at (-.3+1.-.07,-.3-.31){\footnotesize$^{#6}$};
\vertex (X1l) at (1.7,.3);
\vertex (X1lp) at (1.7+.21,.3-.02){\footnotesize$^{#3}$};
\vertex (Xnl) at (1.7,-.3);
\vertex (Xnlp) at (1.7+.21,-.3-.05){\footnotesize$^{#4}$};
   \diagram*{
   (x1r) -- [thick, quarter right,  scalar ,looseness=.8] (X) --[ thick, quarter left, looseness=.8, scalar]  (xnl),
         (x2r) -- [thick, quarter left, looseness=.8, scalar] (X) --[ thick, quarter right, looseness=.8, scalar] (x1l)  ,
  (Xnl)  -- [thick, quarter left, looseness=.8] (Y) -- [thick, quarter right, looseness=.8 , scalar]   (y1l)  ,
  (X1l)   -- [thick, quarter right, looseness=.8] (Y) -- [thick, quarter left, looseness=.8, scalar]    (y2l),
   (c1) -- [very thick, scalar, red] (c2),
};
\node[roundnode] (X1) at (0,0);%{\tiny{$iT$}};
  \end{feynman}
\end{tikzpicture}
  \end{minipage} 
  }

\newcommand{\dCinc}[4]{
  \begin{minipage}[h]{0.12\linewidth}\begin{tikzpicture}
  [
roundnode/.style={circle, draw=black!60, fill=black!6, very thick, %minimum size=7mm,
  inner sep=1pt,
  text width=3mm},
]
\begin{feynman}[small]
\vertex (x1) at (0,0);
\vertex (x1T) at (-.2,0){\footnotesize$^{{#1}}$};
\vertex (x1aux) at (-.05,-.07);
\vertex (x2) at (.72,0);
\vertex (x2T) at (.7+.24,0){\footnotesize$^{{#3}}$};
\vertex (x2aux) at (.75,-.07);
\vertex (x3) at (0,-.7);
\vertex (x1T) at (-.2,-.7){\footnotesize$^{{#4}}$};
\vertex (x3aux) at (-.05,-.7+.07);
\vertex (x4) at (.72,-.7);
\vertex (x1T) at (.7+.24,-.7){\footnotesize$^{{#2}}$};
\vertex (x4aux) at (.7+.05,-.7+.07);
   \diagram*{
   (x1) -- [thick, quarter right, looseness=1.2,dotted] (x2)  ,
   (x1aux) -- [thick, quarter right, looseness=1.2] (x2aux)  ,
   (x3) -- [thick, quarter left, looseness=1.2,dotted] (x4)  , 
   };
  \end{feynman}
\end{tikzpicture}
  \end{minipage} 
  }

\newcommand{\dSis}[4]{
  \begin{minipage}[h]{0.12\linewidth}\begin{tikzpicture}
  [
roundnode/.style={circle, draw=black!60, fill=black!6, very thick, %minimum size=7mm,
  inner sep=1pt,
  text width=3mm},
]
\begin{feynman}[small]
\vertex (x1) at (0,0);
\vertex (x1aux) at (-.05,-.05);
\vertex (x1T) at (-.2,0){\footnotesize$^{{#1}}$};
\vertex (x2) at (.7+.02,0);
\vertex (x2aux) at (.7+.05,-.05);
\vertex (x2T) at (.7+.24,0){\footnotesize$^{{#2}}$};
\vertex (x3) at (0,-.7);
\vertex (x3aux) at (-.05,-.7+.1-.05);
\vertex (x1T) at (-.2,-.7){\footnotesize$^{{#3}}$};
\vertex (x4) at (.7+.02,-.7);
\vertex (x4aux) at (.7+.05,-.7+.1-.05);
\vertex (x1T) at (.7+.24,-.7){\footnotesize$^{{#4}}$};
   \diagram*{
     (x1) -- [thick, quarter right, looseness=1.2, dotted] (x2)  ,
   (x1aux) -- [thick, quarter left, looseness=1.2] (x3aux)  ,
      (x3) -- [thick, quarter left, looseness=1.2, dotted] (x4)  ,
 
   };
  \end{feynman}
\end{tikzpicture}
  \end{minipage} 
  }

\newcommand{\dNMFFun}{
  \begin{minipage}[h]{0.12\linewidth}\begin{tikzpicture}
  [
roundnode/.style={circle, draw=black!60, fill=black!6, very thick, %minimum size=7mm,
  inner sep=1pt,
  text width=3mm},
]
\begin{feynman}[small]
\vertex (X) at (-.3,0);
\vertex (Xaux) at (-.3,-.1);
\vertex (x1l) at (-1.,.4);
\vertex (x1lp) at (-1-.15,.4-.02){\footnotesize$^{4}$};
\vertex (x2l) at (-1.,0);
\vertex (x2lp) at (-1-.15,0){\footnotesize$^{1_i^-}$};
\vertex (xnl) at (-1.,-.4);
\vertex (xnlp) at (-1.-.15,-.4-.02){\footnotesize$^{3_j^-}$};
\vertex (x1r) at (+.5,.3);
\vertex (x2r) at (+.5,-.3);
\vertex (c1) at (.6,.35);
\vertex (c2) at (.6,-.4);
\node [crossed dot] (Y) at (0+1.4,0);
\vertex (y1l) at (-.3+1.,.3);
\vertex (y2l) at (-.3+1.,-.3);
\vertex (X1l) at (2.1,.4);
\vertex (X2l) at (2.1,0);
\vertex (X2lp) at (2.1+.24,0){\footnotesize$^{2^-_A}$};
\vertex (Xnl) at (2.1,-.4);
   \diagram*{
   (x1l) -- [thick, quarter left, looseness=.8, scalar] (X) --[ thick, quarter left, looseness=.8, boson] (x1r) ,
      (x2l) -- [thick ] (X) ,
         (xnl) -- [thick, quarter right, looseness=.8] (X) --[ thick, quarter right, looseness=.8, boson] (x2r) ,
   (y1l) -- [thick, quarter left, looseness=.8, boson] (Y),  %-- [thick, quarter left, looseness=.8] (X1l)   ,
   (y2l) -- [thick, quarter right, looseness=.8, boson] (Y),
   (Y)   -- [thick, boson] (X2l)  ,
   (c1) -- [very thick, scalar, red] (c2),
};
\node[roundnode] (X1) at (-.3,0);%{\tiny{$iT$}};
  \end{feynman}
\end{tikzpicture}
  \end{minipage} 
  }

\newcommand{\dNMFFdos}{
  \begin{minipage}[h]{0.12\linewidth}\begin{tikzpicture}
  [
roundnode/.style={circle, draw=black!60, fill=black!6, very thick, %minimum size=7mm,
  inner sep=1pt,
  text width=3mm},
]
\begin{feynman}[small]
\vertex (X) at (-.3,0);
\vertex (Xaux) at (-.3,-.1);
\vertex (x1l) at (-1.,.4);
\vertex (x1lp) at (-1-.15,.4-.02){\footnotesize$^{4}$};
\vertex (xnl) at (-1.,-.4);
\vertex (xnlp) at (-1.-.15,-.4-.02){\footnotesize$^{1_i^-}$};
\vertex (x1r) at (+.5,.3);
\vertex (x2r) at (+.5,-.3);
\vertex (c1) at (.6,.35);
\vertex (c2) at (.6,-.4);
\node [crossed dot] (Y) at (0+1.3,.3);
\node[dot] (Y2) at (0+1.3,-.3);
\vertex (y1l) at (-.3+1.,.3);
\vertex (y2l) at (-.3+1.,-.3);
\vertex (X1l) at (1.9,.4);
\vertex (X2l) at (1.9,.3);
\vertex (X2lp) at (1.9+.24,.3){\footnotesize$^{2^-_A}$};
\vertex (Xnl) at (1.9,-.3);
\vertex (Xnlp) at (1.9+.24,-.2-.08){\footnotesize$^{3_j^-}$};
   \diagram*{
   (x1l) -- [thick, quarter left, looseness=.8, scalar] (X) --[ thick, quarter left, looseness=.8, boson] (x1r) ,
         (xnl) -- [thick, quarter right, looseness=.8] (X) --[ thick, quarter right, looseness=.8 ] (x2r) ,
   (y1l) -- [thick,  boson] (Y),  %-- [thick, quarter left, looseness=.8] (X1l)   ,
   (y2l) -- [thick ] (Y2) -- [thick, boson] (Y),
   (Y)   -- [thick,  boson] (X2l)  ,
    (Y2)   -- [thick ] (Xnl)  ,
   (c1) -- [very thick, scalar, red] (c2),
};
\node[roundnode] (X1) at (-.3,0);%{\tiny{$iT$}};
  \end{feynman}
\end{tikzpicture}
  \end{minipage} 
  }

\newcommand{\dNMFFtres}{
  \begin{minipage}[h]{0.12\linewidth}\begin{tikzpicture}
  [
roundnode/.style={circle, draw=black!60, fill=black!6, very thick, %minimum size=7mm,
  inner sep=1pt,
  text width=3mm},
]
\begin{feynman}[small]
\vertex (X) at (-.3,0);
\vertex (Xaux) at (-.3,-.1);
\vertex (x1l) at (-1.,.4);
\vertex (x1lp) at (-1-.15,.4-.02){\footnotesize$^{4}$};
\vertex (xnl) at (-1.,-.4);
\vertex (xnlp) at (-1.-.15,-.4-.02){\footnotesize$^{3_j^-}$};
\vertex (x1r) at (+.5,.3);
\vertex (x2r) at (+.5,-.3);
\vertex (c1) at (.6,.35);
\vertex (c2) at (.6,-.4);
\node [crossed dot] (Y) at (0+1.3,.3);
\node[dot] (Y2) at (0+1.3,-.3);
\vertex (y1l) at (-.3+1.,.3);
\vertex (y2l) at (-.3+1.,-.3);
\vertex (X1l) at (1.9,.4);
\vertex (X2l) at (1.9,.3);
\vertex (X2lp) at (1.9+.24,.3){\footnotesize$^{2^-_A}$};
\vertex (Xnl) at (1.9,-.3);
\vertex (Xnlp) at (1.9+.24,-.2-.08){\footnotesize$^{1_i^-}$};
   \diagram*{
   (x1l) -- [thick, quarter left, looseness=.8, scalar] (X) --[ thick, quarter left, looseness=.8, boson] (x1r) ,
         (xnl) -- [thick, quarter right, looseness=.8] (X) --[ thick, quarter right, looseness=.8 ] (x2r) ,
   (y1l) -- [thick,  boson] (Y),  %-- [thick, quarter left, looseness=.8] (X1l)   ,
   (y2l) -- [thick ] (Y2) -- [thick, boson] (Y),
   (Y)   -- [thick,  boson] (X2l)  ,
    (Y2)   -- [thick ] (Xnl)  ,
   (c1) -- [very thick, scalar, red] (c2),
};
\node[roundnode] (X1) at (-.3,0);%{\tiny{$iT$}};
  \end{feynman}
\end{tikzpicture}
  \end{minipage} 
  }

\newcommand{\dOytoOdDos}{
  \begin{minipage}[h]{0.12\linewidth}\begin{tikzpicture}
  [
roundnode/.style={circle, draw=black!60, fill=black!6, very thick, %minimum size=7mm,
  inner sep=1pt,
  text width=3mm},
]
\begin{feynman}[small]
\vertex (X) at (-.3,0);
\vertex (Xaux) at (-.3,-.1);
\vertex (x1l) at (-1.,.4);
\vertex (x1lp) at (-1-.2,.4){\footnotesize$^{4^-}$};
\vertex (xnl) at (-1.,-.4);
\vertex (xnlp) at (-1.-.15,-.4){\footnotesize$^{1^-_i}$};
\vertex (x1r) at (+.3,.3);
\vertex (x2r) at (+.3,-.3);
\vertex (x3r) at (+.3,0);
\vertex (c1) at (.4,.35);
\vertex (c2) at (.4,-.4);
\node [crossed dot] (Y) at (0+1.1,0);
\vertex (y1l) at (-.5+1.,.3);
\vertex (y2l) at (-.5+1.,-.3);
\vertex (y3l) at (-.5+1.,0);
\vertex (X1l) at (1.8,.4);
\vertex (X1lp) at (1.8+.16,.35){\footnotesize$^{3_j}$};
\vertex (Xnl) at (1.8,-.4);
\vertex (Xnlp) at (1.8+.22,-.4){\footnotesize$^{2^-}$};
   \diagram*{
   (x1l) -- [thick, quarter left, looseness=.8, boson] (X) --[ thick, quarter left, looseness=.8, scalar] (x1r) ,
      (x3r) -- [thick, scalar] (X) ,
         (xnl) -- [thick, quarter right, looseness=.8] (X) --[ thick, quarter right, looseness=.8] (x2r) ,
   (y1l) -- [thick, quarter left, looseness=.8, scalar] (Y), 
     (y3l) -- [thick, scalar] (Y), 
        (y2l) -- [thick, quarter right, looseness=.8] (Y),
   (Y)   -- [thick, quarter left, looseness=.8, scalar] (X1l)  ,
    (Y)   -- [thick, quarter right, looseness=.8] (Xnl)  ,
   (c1) -- [very thick, scalar, red] (c2),
};
\node[roundnode] (X1) at (-.3,0);%{\tiny{$iT$}};
  \end{feynman}
\end{tikzpicture}
  \end{minipage} 
  }

\newcommand{\dOytoOpsipsi}{
  \begin{minipage}[h]{0.12\linewidth}\begin{tikzpicture}
  [
roundnode/.style={circle, draw=black!60, fill=black!6, very thick, %minimum size=7mm,
  inner sep=1pt,
  text width=3mm},
]
\begin{feynman}[small]
\vertex (X) at (-.3,0);
\vertex (Xaux) at (-.3,-.1);
\vertex (x1l) at (-1.,.4);
\vertex (x1lp) at (-1-.2,.4){\footnotesize$^{\l^-}$};
\vertex (xnl) at (-1.,-.4);
\vertex (xnlp) at (-1.-.2,-.4){\footnotesize$^{e^-}$};
\vertex (x1r) at (+.3,.3);
\vertex (x2r) at (+.3,-.3);
\vertex (x3r) at (+.3,0);
\vertex (c1) at (.4,.35);
\vertex (c2) at (.4,-.4);
\node [crossed dot] (Y) at (0+1.1,0);
\vertex (y1l) at (-.5+1.,.3);
\vertex (y2l) at (-.5+1.,-.3);
\vertex (y3l) at (-.5+1.,0);
\vertex (X1l) at (1.8,.4);
\vertex (X1lp) at (1.85+.2,.43){\footnotesize$^{q_i^-}$};
\vertex (Xnl) at (1.8,-.4);
\vertex (Xnlp) at (1.85+.2,-.4){\footnotesize$^{u^-}$};
   \diagram*{
   (x1l) -- [thick, quarter left, looseness=.8] (X) --[ thick, quarter left, looseness=.8, scalar] (x1r) ,
      (x3r) -- [thick, scalar] (X) ,
         (xnl) -- [thick, quarter right, looseness=.8] (X) --[ thick, quarter right, looseness=.8, scalar] (x2r) ,
   (y1l) -- [thick, quarter left, looseness=.8, scalar] (Y), 
     (y3l) -- [thick, scalar] (Y), 
        (y2l) -- [thick, quarter right, looseness=.8, scalar] (Y),
   (Y)   -- [thick, quarter left, looseness=.8] (X1l)  ,
    (Y)   -- [thick, quarter right, looseness=.8] (Xnl)  ,
   (c1) -- [very thick, scalar, red] (c2),
};
\node[roundnode] (X1) at (-.3,0);%{\tiny{$iT$}};
  \end{feynman}
\end{tikzpicture}
  \end{minipage} 
  }

\newcommand{\dOsixtoOy}{
  \begin{minipage}[h]{0.12\linewidth}\begin{tikzpicture}
  [
roundnode/.style={circle, draw=black!60, fill=black!6, very thick, %minimum size=7mm,
  inner sep=1pt,
  text width=3mm},
]
\begin{feynman}[small]
\vertex (X) at (-.3,0);
\vertex (Xaux) at (-.3,-.1);
\vertex (x1l) at (-1.,.4);
\vertex (x1lp) at (-1-.2,.4){\footnotesize$^{-}$};
\vertex (xnl) at (-1.,-.4);
\vertex (xnlp) at (-1.-.2,-.4){\footnotesize$^{-}$};
\vertex (x1r) at (+.3,.3);
\vertex (x2r) at (+.3,-.3);
\vertex (x3r) at (+.3,0);
\vertex (c1) at (.4,.35);
\vertex (c2) at (.4,-.4);
\node [crossed dot] (Y) at (0+1.1,0);
\vertex (y1l) at (-.5+1.,.3);
\vertex (y2l) at (-.5+1.,-.3);
\vertex (y3l) at (-.5+1.,0);
\vertex (X1l) at (1.8,.4);
\vertex (X1lp) at (1.85+.2,.43);%{\footnotesize$^{q_i^-}$};
\vertex (Xnl) at (1.8,-.4);
\vertex (X2l) at (1.8,0);
\vertex (Xnlp) at (1.85+.2,-.4);%{\footnotesize$^{u^-}$};
   \diagram*{
   (x1l) -- [thick, quarter left, looseness=.8] (X) --[ thick, quarter left, looseness=.8, scalar] (x1r) ,
      (x3r) -- [thick, scalar] (X) ,
         (xnl) -- [thick, quarter right, looseness=.8] (X) --[ thick, quarter right, looseness=.8, scalar] (x2r) ,
   (y1l) -- [thick, quarter left, looseness=.8, scalar] (Y), 
     (y3l) -- [thick, scalar] (Y), 
        (y2l) -- [thick, quarter right, looseness=.8, scalar] (Y),
   (Y)   -- [thick, quarter left, looseness=.8, scalar] (X1l)  ,
   (Y)   -- [thick,  scalar] (X2l)  ,
    (Y)   -- [thick, quarter right, looseness=.8, scalar] (Xnl)  ,
   (c1) -- [very thick, scalar, red] (c2),
};
\node[roundnode] (X1) at (-.3,0);%{\tiny{$iT$}};
  \end{feynman}
\end{tikzpicture}
  \end{minipage} 
  }

\newcommand{\twoloopgenUn}{
  \begin{minipage}[h]{0.12\linewidth}\begin{tikzpicture}
  [
roundnode/.style={circle, draw=black!60, fill=black!6, very thick, %minimum size=7mm,
  inner sep=1.6pt,
  text width=3mm},
]
\begin{feynman}[small]
\vertex (X) at (-.3,0);
\vertex (Xaux) at (-.3,-.1);
\vertex (x1l) at (-1.,.4);
\vertex (xnl) at (-1.,-.4);
\vertex (x1r) at (+.3,.3);
\vertex (x2r) at (+.3,-.3);
\vertex (x3r) at (+.3,0);
\vertex (c1) at (.4,.35);
\vertex (c2) at (.4,-.4);
\node [crossed dot] (Y) at (0+1.1+.1,0);
\vertex (Yaux) at (1.08+.1,-.09);
\vertex (y1l) at (-.5+1.,.3);
\vertex (y2l) at (-.5+1.,-.3);
\vertex (y3l) at (-.5+1.,0);
\vertex (X1l) at (1.8+.2,.4);
\vertex (X2l) at (1.8+.2,.2);
\vertex (Xnl) at (1.8+.2,-.4);
   \diagram*{
   (x1l) -- [thick, quarter left, looseness=.8] (X) --[ thick, quarter left, looseness=.8] (x1r) ,
      (x3r) -- [thick] (X) ,
         (xnl) -- [thick, quarter right, looseness=.8] (X) --[ thick, quarter right, looseness=.8] (x2r) ,
   (y1l) -- [thick, quarter left, looseness=.8] (Y), 
     (y3l) -- [thick] (Y), 
        (y2l) -- [thick, quarter right, looseness=.8] (Y),
   (Yaux)   -- [thick, quarter left, looseness=.8] (X2l)  ,
   (Y)   -- [thick, quarter left, looseness=.8] (X1l)  ,
    (Y)   -- [thick, quarter right, looseness=.8] (Xnl)  ,
   (c1) -- [very thick, scalar, red] (c2),
};
\vertex (D1) at (1.6+.2,.03){$\vdots$};
\node[roundnode] (Y1) at (0+1.1+.1,0){\tiny{$\hspace{-.1cm}\vspace{-.05cm}O_i^{(0)}$}};
\node[roundnode] (X1) at (-.3,0){\tiny{$iT$}};
  \end{feynman}
\end{tikzpicture}
  \end{minipage} 
  }

\newcommand{\twoloopgenDos}{
  \begin{minipage}[h]{0.12\linewidth}\begin{tikzpicture}
  [
roundnode/.style={circle, draw=black!60, fill=black!6, very thick, %minimum size=7mm,
  inner sep=1.6pt,
  text width=3mm},
]
\begin{feynman}[small]
\vertex (X) at (-.3,0);
\vertex (Xaux) at (-.3,-.1);
\vertex (x1l) at (-1.,.4);
\vertex (xnl) at (-1.,-.4);
\vertex (x1r) at (+.3,.3);
\vertex (x2r) at (+.3,-.3);
\vertex (c1) at (.4,.35);
\vertex (c2) at (.4,-.4);
\node [crossed dot] (Y) at (0+1.1+.1,0);
\vertex (Yaux) at (1.08+.1,-.09);
\vertex (y1l) at (-.5+1.,.3);
\vertex (y2l) at (-.5+1.,-.3);
\vertex (X1l) at (1.8+.2,.4);
\vertex (X2l) at (1.8+.2,.2);
\vertex (Xnl) at (1.8+.2,-.4);
   \diagram*{
   (x1l) -- [thick, quarter left, looseness=.8] (X) --[ thick, quarter left, looseness=.8] (x1r) ,
         (xnl) -- [thick, quarter right, looseness=.8] (X) --[ thick, quarter right, looseness=.8] (x2r) ,
   (y1l) -- [thick, quarter left, looseness=.8] (Y), 
        (y2l) -- [thick, quarter right, looseness=.8] (Y),
   (Yaux)   -- [thick, quarter left, looseness=.8] (X2l)  ,
   (Y)   -- [thick, quarter left, looseness=.8] (X1l)  ,
    (Y)   -- [thick, quarter right, looseness=.8] (Xnl)  ,
   (c1) -- [very thick, scalar, red] (c2),
};
\vertex (D1) at (1.6+.2,.03){$\vdots$};
\node[roundnode] (Y1) at (0+1.1+.1,0){\tiny{$\hspace{-.1cm}\vspace{-.05cm}O_i^{(1)}$}};
\node[roundnode] (X1) at (-.3,0){\tiny{$iT$}};
  \end{feynman}
\end{tikzpicture}
  \end{minipage} 
  }

\newcommand{\dMaster}{
  \begin{minipage}[h]{0.12\linewidth}\begin{tikzpicture}
  [
roundnode/.style={circle, draw=black!60, fill=black!6, very thick, %minimum size=7mm,
  inner sep=1pt,
  text width=3mm},
  squarednode/.style={rectangle, draw=gray!50, fill=gray!10, very thick, minimum size=5mm},
  squarednode2/.style={rectangle, draw=white!100, fill=white!100,   inner sep=2pt,
  text width=7.5mm}
]
\begin{feynman}[small]
\vertex (v1) at (-.2,1);
\vertex (v2) at (5,1);
\vertex (v3) at (9.7,1);
\vertex (x11) at (0,.15);
\vertex  (x12) at (0,0);
\vertex (x21) at (2+.5,.15);
\vertex (x211) at (3.8+.5,.15);
\vertex (x22) at (2+.3,0);
\vertex (x221) at (3.8+.4,0);
\vertex  (x31) at (6.5+1,.15);
\vertex (x32) at (6.5+.9,0);
\vertex  (x41) at (8+1.5,.15);
\vertex  (x42) at (8+1.5,0);
\vertex (t1) at (.03,.65){$^{\langle \  \cdot  \  \rangle^3}$};
\vertex (t2) at (3.3+.3,.65){$^{\langle \  \cdot   \  \rangle^2}$};
\vertex  (t3) at (6.3+.93,.65){$^{\langle \  \cdot   \  \rangle}$};
\vertex  (t4) at (8+1.5,.55){$^{1}$};
\vertex  (t5) at (3.5,-1.9){$^{\langle \    \cdot  \  \rangle [ \  \cdot   \   ]}$};
\vertex (t6) at (1.15,-.4){$\scriptsize{^{\text{2-loop}}}$};
\vertex (t7) at (1.65,+1.2){$\scriptsize{^{\text{1-loop}}}$};
\vertex (t9) at (1.2,.26){\textbf{$^\text{\ref{case1}}$}};
\vertex (t10) at (5.9,.26){\textbf{$^\text{\ref{case2}}$}};
\vertex (t11) at (8.5,.26){\textbf{$^\text{\ref{case3}}$}};
   \diagram*{
   (x42) -- [very thick, fermion, darkblue] (x32)-- [very thick, fermion, darkblue] (x221) -- [very thick, fermion, darkblue] (x22) -- [very thick, fermion, darkblue] (x12) ,
   };
   \node[squarednode] (X1) at (0,0){$F^3$};
   \node[squarednode] (X2) at (3+.5,0){ $\phi^2 F^2\ \, \psi^2 \phi F\  \, \psi^4$ };
   \node[squarednode] (X2p) at (3.5,-1.2){ $\partial^2 \phi^2 \ \, \bar\psi \psi  \partial \phi^2 \  \, \bar\psi^2 \psi^2$ };
   \node[squarednode] (X3) at (6.2+1,0){ $\psi^2 \phi^3$ };
   \node[squarednode] (X4) at (8+1.5,0){ $\phi^6$ };
   \vertex (A) at (0,-.5);
   \vertex (B) at (.5,-1);
   \vertex (C) at (1.4,-1.2);
   \draw[->, to path={-| (\tikztotarget)}, thick, rounded corners] (A) -- (B) -- (C);
   \node[squarednode2] (t8) at (.3,-1){$\scriptsize{^{\cancel{\text{1-loop}}}}$};
   \vertex (D) at (4.8,-.2);
   \vertex (E) at (5.,-.6);
   \vertex (F) at (4.8,-.95);
   \draw[<->, to path={-| (\tikztotarget)}, thick, rounded corners] (D) -- (E) -- (F);
   \vertex (V1) at (-.1,1);
   \vertex (V2) at (2.5,1);
   \vertex (V3) at (2.8,1);
   \vertex (V4) at (9.6,1);
   \draw[->, to path={-| (\tikztotarget)}, thick, rounded corners] (V1) -- (V2);
   \draw[->, to path={-| (\tikztotarget)}, thick, rounded corners] (V2) -- (V3);
   \draw[->, to path={-| (\tikztotarget)}, thick, rounded corners] (V3) -- (V4);
  \end{feynman}
\end{tikzpicture}
  \end{minipage} 
  }

\newcommand{\dBCFWun}{
  \begin{minipage}[h]{0.12\linewidth}\begin{tikzpicture}
  [
roundnode/.style={circle, draw=black!60, fill=black!6, very thick, %minimum size=7mm,
  inner sep=1pt,
  text width=3mm},
]
\begin{feynman}[small]
\vertex (X) at (0,0);
\vertex (Xaux) at (0,-.1);
\vertex (x1l) at (-.7,.4);
\vertex (x1lp) at (-.7-.15,.4-.02){\footnotesize$^{4}$};
\vertex (xnl) at (-.7,-.4);
\vertex (xnlp) at (-.7-.15,-.4-.05){\footnotesize$^{\hat{1}_i^-}$};
\vertex (x1r) at (+.5,.3);
\vertex (x2r) at (+.5,-.3);
\vertex (Y) at (0+1.2,0);
\vertex (Yaux) at (1.2,-.1);
\vertex (y1l) at (-.3+1.,.3);
\vertex (y2l) at (-.3+1.,-.3);
\vertex (X1l) at (1.9,.4);
\vertex (X1lp) at (1.9+.24,.4-.02){\footnotesize$^{3_j^-}$};
\vertex (X2l) at (1.9,.2);
\vertex (Xnl) at (1.9,-.4);
\vertex (Xnlp) at (1.9+.24,-.4-.05){\footnotesize$^{\hat{2}_A^{\pm}}$};
   \diagram*{
   (x1l) -- [thick, quarter left, looseness=.8, scalar] (X) --[ thick] (Y) ,
         (xnl) -- [thick, quarter right, looseness=.8] (X), % --[ thick, quarter right, looseness=.8] (x2r) ,
   (Y) -- [thick, quarter left, looseness=.8] (X1l)   ,
    (Y)   -- [thick, quarter right, looseness=.8, boson] (Xnl)  ,
};
\node[roundnode] (Y1) at (0+1.2,0){\tiny{$\, R$}};
\node[roundnode] (X1) at (0,0){\tiny{$\, L$}};
  \end{feynman}
\end{tikzpicture}
  \end{minipage} 
  }

\newcommand{\dBCFWdos}{
  \begin{minipage}[h]{0.12\linewidth}\begin{tikzpicture}
  [
roundnode/.style={circle, draw=black!60, fill=black!6, very thick, %minimum size=7mm,
  inner sep=1pt,
  text width=3mm},
]
\begin{feynman}[small]
\vertex (X) at (0,0);
\vertex (Xaux) at (0,-.1);
\vertex (x1l) at (-.7,.4);
\vertex (x1lp) at (-.7-.2,.43){\footnotesize$^{\hat{2}_A^+}$};
\vertex (xnl) at (-.7,-.4);
\vertex (xnlp) at (-.7-.16,-.4-.02){\footnotesize$^{1_i^-}$};
\vertex (x1r) at (+.5,.3);
\vertex (x2r) at (+.5,-.3);
\vertex (Y) at (0+1.2,0);
\vertex (Yaux) at (1.2,-.1);
\vertex (y1l) at (-.3+1.,.3);
\vertex (y2l) at (-.3+1.,-.3);
\vertex (X1l) at (1.9,.4);
\vertex (X1lp) at (1.9+.27,.43){\footnotesize$^{\hat{3}_A^+}$};
\vertex (X2l) at (2,.0);
\vertex (X2lp) at (1.9+.24,.0-.07){\footnotesize$^{5}$};
\vertex (Xnl) at (1.9,-.4);
\vertex (Xnlp) at (1.9+.24,-.4-.05){\footnotesize$^{4_j^{-}}$};
   \diagram*{
   (x1l) -- [thick, quarter left, looseness=.8, boson] (X) --[ thick] (Y) ,
         (xnl) -- [thick, quarter right, looseness=.8] (X), % --[ thick, quarter right, looseness=.8] (x2r) ,
   (Y) -- [thick, quarter left, looseness=.8 , boson] (X1l)   ,
   (Y)   -- [thick,  scalar] (X2l)  ,
    (Y)   -- [thick, quarter right, looseness=.8 ] (Xnl)  ,
};
\node[roundnode] (Y1) at (0+1.2,0){\tiny{$\, R$}};
\node[roundnode] (X1) at (0,0){\tiny{$\, L$}};
  \end{feynman}
\end{tikzpicture}
  \end{minipage} 
  }

\newcommand{\dBCFWtres}{
  \begin{minipage}[h]{0.12\linewidth}\begin{tikzpicture}
  [
roundnode/.style={circle, draw=black!60, fill=black!6, very thick, %minimum size=7mm,
  inner sep=1pt,
  text width=3mm},
]
\begin{feynman}[small]
\vertex (X) at (0,0);
\vertex (Xaux) at (0,-.1);
\vertex (x1l) at (-.7,.4);
\vertex (x1lp) at (-.7-.2,.43){\footnotesize$^{\hat{2}_A^+}$};
\vertex (x2l) at (-.7,0);
\vertex (x2lp) at (-.7-.2,-.07){\footnotesize$^{5}$};
\vertex (xnl) at (-.7,-.4);
\vertex (xnlp) at (-.7-.16,-.4-.02){\footnotesize$^{1_i^-}$};
\vertex (x1r) at (+.5,.3);
\vertex (x2r) at (+.5,-.3);
\vertex (Y) at (0+1.2,0);
\vertex (Yaux) at (1.2,-.1);
\vertex (y1l) at (-.3+1.,.3);
\vertex (y2l) at (-.3+1.,-.3);
\vertex (X1l) at (1.9,.4);
\vertex (X1lp) at (1.9+.24,.43){\footnotesize$^{\hat{3}_A^+}$};
\vertex (X2l) at (1.9,.2);
\vertex (Xnl) at (1.9,-.4);
\vertex (Xnlp) at (1.9+.24,-.4-.05){\footnotesize$^{4_j^{-}}$};
   \diagram*{
   (x1l) -- [thick, quarter left, looseness=.8, boson] (X) --[ thick] (Y) ,
     (x2l) -- [thick ,  scalar] (X) ,
         (xnl) -- [thick, quarter right, looseness=.8] (X), % --[ thick, quarter right, looseness=.8] (x2r) ,
   (Y) -- [thick, quarter left, looseness=.8 , boson] (X1l)   ,
    (Y)   -- [thick, quarter right, looseness=.8 ] (Xnl)  ,
};
\node[roundnode] (Y1) at (0+1.2,0){\tiny{$\, R$}};
\node[roundnode] (X1) at (0,0){\tiny{$\, L$}};
  \end{feynman}
\end{tikzpicture}
  \end{minipage} 
  }

\newcommand\dgenUn[9]{%
    \def\tempa{#1}%
    \def\tempb{#2}%
    \def\tempc{#3}%
    \def\tempd{#4}%
    \def\tempe{#5}%
    \def\tempf{#6}%
    \def\tempg{#7}%
    \def\temph{#8}%
    \def\tempi{#9}%
    \dgenUnCont
}

\newcommand\dgenUnCont[1]{
  \begin{minipage}[h]{0.12\linewidth}\begin{tikzpicture}
  [
roundnode/.style={circle, draw=black!60, fill=black!6, very thick, %minimum size=7mm,
  inner sep=1pt,
  text width=3mm},
]
\begin{feynman}[small]
\vertex (X) at (-.3,0);
\vertex (Xaux) at (-.3,-.1);
\vertex (x1l) at (-1.,.4);
\vertex (xnl) at (-1.,-.4);
\vertex (x1r) at (+.3,.3);
\vertex (x2r) at (+.3,-.3);
\vertex (x3r) at (+.3,0);
\vertex (c1) at (.4,.35);
\vertex (c2) at (.4,-.4);
\node [crossed dot] (Y) at (0+1.1,0);
\vertex (Yaux) at (1.08,-.09);
\vertex (y1l) at (-.5+1.,.3);
\vertex (y2l) at (-.5+1.,-.3);
\vertex (y3l) at (-.5+1.,0);
\vertex (X1l) at (1.8,.4);
\vertex (Xnl) at (1.8,-.4);
   \diagram*{
   (x1l) -- [thick, quarter left, looseness=.8, \tempa] (X) --[ thick, quarter left, looseness=.8, \tempc] (x1r) ,
      (x3r) -- [thick, \tempd] (X) ,
         (xnl) -- [thick, quarter right, looseness=.8, \tempb] (X) --[ thick, quarter right, looseness=.8, \tempe] (x2r) ,
   (y1l) -- [thick, quarter left, looseness=.8, \tempf] (Y), 
     (y3l) -- [thick, \tempg] (Y), 
        (y2l) -- [thick, quarter right, looseness=.8, \temph] (Y),
   (Y)   -- [thick, quarter left, looseness=.8, \tempi] (X1l)  ,
    (Y)   -- [thick, quarter right, looseness=.8,  #1] (Xnl)  ,
   (c1) -- [very thick, scalar, red] (c2),
};
\node[roundnode] (X1) at (-.3,0);%{\tiny{$iT$}};
  \end{feynman}
\end{tikzpicture}
  \end{minipage} 
  }

\newcommand\dgenDos[9]{%
    \def\tempa{#1}%
    \def\tempb{#2}%
    \def\tempc{#3}%
    \def\tempd{#4}%
    \def\tempe{#5}%
    \def\tempf{#6}%
    \def\tempg{#7}%
    \def\temph{#8}%
    \def\tempi{#9}%
    \dgenDosCont
}
\newcommand\dgenDosCont[1]{
  \begin{minipage}[h]{0.12\linewidth}\begin{tikzpicture}
  [
roundnode/.style={circle, draw=black!60, fill=black!6, very thick, %minimum size=7mm,
  inner sep=1pt,
  text width=3mm},
]
\begin{feynman}[small]
\vertex (X) at (-.3,0);
\vertex (Xaux) at (-.3,-.1);
\vertex (x1l) at (-1.,.4);
\vertex (xnl) at (-1.,-.4);
\vertex (x1r) at (+.3,.3);
\vertex (x2r) at (+.3,-.3);
\vertex (c1) at (.4,.35);
\vertex (c2) at (.4,-.4);
\node [crossed dot] (Y) at (0+1.1+.1,0);
\vertex (Yaux) at (1.08+.1,-.09);
\vertex (y1l) at (-.5+1.,.3);
\vertex (y2l) at (-.5+1.,-.3);
\vertex (y3l) at (-0.25+1.05,-.2);
\vertex (X1l) at (1.8+.1,.4);
\vertex (Xnl) at (1.8+.1,-.4);
   \diagram*{
   (x1l) -- [thick, quarter left, looseness=.8, \tempa] (X) --[ thick, quarter left, looseness=.8, \tempc] (x1r) ,
         (xnl) -- [thick, quarter right, looseness=.8, \tempb] (X) --[ thick, quarter right, looseness=.8, \tempd] (x2r) ,
   (y1l) -- [thick, quarter left, looseness=.8, \tempe] (Y), 
     (y3l) -- [thick,  quarter left, looseness=1.2, \tempf] (Y), 
    (Y) --  [thick, quarter left, looseness=1.2, \tempg]   (y3l) -- [thick,\temph ] (y2l) ,
   (Y)   -- [thick, quarter left, looseness=.8, \tempi] (X1l)  ,
    (Y)   -- [thick, quarter right, looseness=.8, #1] (Xnl)  ,
   (c1) -- [very thick, scalar, red] (c2),
};
\node[roundnode] (X1) at (-.3,0);%{\tiny{$iT$}};
  \end{feynman}
\end{tikzpicture}
  \end{minipage} 
  }

\newcommand\dgenUnP[9]{
  \begin{minipage}[h]{0.12\linewidth}\begin{tikzpicture}
  [
roundnode/.style={circle, draw=black!60, fill=black!6, very thick, %minimum size=7mm,
  inner sep=1pt,
  text width=3mm},
]
\begin{feynman}[small]
\vertex (X) at (-.3,0);
\vertex (Xaux) at (-.3,-.1);
\vertex (x1l) at (-1.,.4);
\vertex (x1lT) at (-1.2,.4){\footnotesize$^{-}$};
\vertex (xnl) at (-1.,-.4);
\vertex (xnlT) at (-1.2,-.4){\footnotesize$^{-}$};
\vertex (x1r) at (+.3,.3);
\vertex (x2r) at (+.3,-.3);
\vertex (x3r) at (+.3,0);
\vertex (c1) at (.4,.35);
\vertex (c2) at (.4,-.4);
\node [crossed dot] (Y) at (0+1.1,0);
\vertex (Yaux) at (1.08,-.09);
\vertex (y1l) at (-.5+1.,.3);
\vertex (y2l) at (-.5+1.,-.3);
\vertex (y3l) at (-.5+1.,0);
\vertex (X1l) at (1.8,0);
\vertex (X1lT) at (2,0){\footnotesize$^{-}$};
   \diagram*{
   (x1l) -- [thick, quarter left, looseness=.8, #1] (X) --[ thick, quarter left, looseness=.8, #3] (x1r) ,
      (x3r) -- [thick, #4] (X) ,
         (xnl) -- [thick, quarter right, looseness=.8,  #1] (X) --[ thick, quarter right, looseness=.8, #5] (x2r) ,
   (y1l) -- [thick, quarter left, looseness=.8,  #6] (Y), 
     (y3l) -- [thick, #7] (Y), 
        (y2l) -- [thick, quarter right, looseness=.8,  #8] (Y),
   (Y)   -- [thick,  #9] (X1l)  ,
   (c1) -- [very thick, scalar, red] (c2),
};
\node[roundnode] (X1) at (-.3,0);%{\tiny{$iT$}};
  \end{feynman}
\end{tikzpicture}
  \end{minipage} 
  }

\newcommand\dgenDosP[9]{
  \begin{minipage}[h]{0.12\linewidth}\begin{tikzpicture}
  [
roundnode/.style={circle, draw=black!60, fill=black!6, very thick, %minimum size=7mm,
  inner sep=1pt,
  text width=3mm},
]
\begin{feynman}[small]
\vertex (X) at (-.3,0);
\vertex (Xaux) at (-.3,-.1);
\vertex (x1l) at (-1.,.4);
\vertex (x1lT) at (-1.2,.4){\footnotesize$^{-}$};
\vertex (xnl) at (-1.,-.4);
\vertex (xnlT) at (-1.2,-.4){\footnotesize$^{-}$};
\vertex (x1r) at (+.3,.3);
\vertex (x2r) at (+.3,-.3);
\vertex (c1) at (.4,.35);
\vertex (c2) at (.4,-.4);
\node [crossed dot] (Y) at (0+1.1+.1,0);
\vertex (Yaux) at (1.08+.1,-.09);
\vertex (y1l) at (-.5+1.,.3);
\vertex (y2l) at (-.5+1.,-.3);
\vertex (y3l) at (-0.25+1.05,-.2);
\vertex (X1l) at (1.8+.1,0);
\vertex (X1lT) at (2.1,0){\footnotesize$^{-}$};
   \diagram*{
   (x1l) -- [thick, quarter left, looseness=.8, #1] (X) --[ thick, quarter left, looseness=.8, #3] (x1r) ,
         (xnl) -- [thick, quarter right, looseness=.8, #2] (X) --[ thick, quarter right, looseness=.8,  #4] (x2r) ,
   (y1l) -- [thick, quarter left, looseness=.8,  #5] (Y), 
     (y3l) -- [thick,  quarter left, looseness=1.2, #6] (Y), 
    (Y) --  [thick, quarter left, looseness=1.2, #7]   (y3l) -- [thick, #8] (y2l) ,
   (Y)   -- [thick,   #9] (X1l)  ,
   (c1) -- [very thick, scalar, red] (c2),
};
\node[roundnode] (X1) at (-.3,0);%{\tiny{$iT$}};
  \end{feynman}
\end{tikzpicture}
  \end{minipage} 
  }

\newcommand\dgenDosPP[9]{
  \begin{minipage}[h]{0.12\linewidth}\begin{tikzpicture}
  [
roundnode/.style={circle, draw=black!60, fill=black!6, very thick, %minimum size=7mm,
  inner sep=1pt,
  text width=3mm},
]
\begin{feynman}[small]
\vertex (X) at (-.3,0);
\vertex (Xaux) at (-.3,-.1);
\vertex (x1l) at (-1.,.4);
\vertex (x1lT) at (-1.2,.4){\footnotesize$^{-}$};
\vertex (xnl) at (-1.,-.4);
\vertex (xnlT) at (-1.2,-.4){\footnotesize$^{-}$};
\vertex (x1r) at (+.3,.3);
\vertex (x2r) at (+.3,-.3);
\vertex (c1) at (.4,.35);
\vertex (c2) at (.4,-.4);
\node [crossed dot] (Y) at (0+.9,0);
\vertex (y1l) at (-.5+1.,.3);
\vertex (y2l) at (-.5+1.,-.3);
\vertex (y3l) at (1.5,0);
\vertex (X1l) at (1.8+.1,0);
\vertex (X1lT) at (2.1,0){\footnotesize$^{-}$};
   \diagram*{
   (x1l) -- [thick, quarter left, looseness=.8, #1] (X) --[ thick, quarter left, looseness=.8, #3] (x1r) ,
         (xnl) -- [thick, quarter right, looseness=.8, #2] (X) --[ thick, quarter right, looseness=.8,  #4] (x2r) ,
   (y1l) -- [thick, quarter left, looseness=.8,  #5] (Y), 
     (y3l) -- [thick,  quarter left, looseness=1.2, #6] (Y) -- [thick,  quarter left, looseness=1.2, #7] (y3l), 
    (Y) --  [thick, quarter left, looseness=.8, #8]    (y2l) ,
   (y3l)   -- [thick,   #9] (X1l)  ,
   (c1) -- [very thick, scalar, red] (c2),
};
\node[roundnode] (X1) at (-.3,0);%{\tiny{$iT$}};
  \end{feynman}
\end{tikzpicture}
  \end{minipage} 
  }

\usepackage{ragged2e}
\usepackage{array}
\newcolumntype{L}[1]{>{\raggedright\let\newline\\\arraybackslash\hspace{0pt}}m{#1}}
\newcolumntype{C}[1]{>{\centering\let\newline\\\arraybackslash\hspace{0pt}}m{#1}}
\newcolumntype{R}[1]{>{\raggedleft\let\newline\\\arraybackslash\hspace{0pt}}m{#1}}

\usepackage{dsfont} 
\usepackage{tcolorbox}
\usepackage{booktabs}

\usepackage{titlesec}
\titleformat*{\section}{\large\bfseries}
\titleformat*{\subsection}{\normalsize\bfseries}
\titleformat*{\subsubsection}{\normalsize\it}
\titleformat*{\paragraph}{\normalsize\bfseries}
\titleformat*{\subparagraph}{\normalsize\bfseries}

\def\a{\alpha}
\def\b{\beta}

\def\lra#1{\overset{\text{\scriptsize$\leftrightarrow$}}{#1}}

\newcommand{\reef}[1]{(\ref{#1})}

\def\eps{\epsilon}

\newcommand{\beq}{\begin{equation}} 
\newcommand{\eeq}{\end{equation}}
 
\def\nn{\nonumber}

\def\le{\leqslant}
\def\geq{\geqslant}

\newcommand{\diffop}[2]{\ifthenelse{\equal{#2}{1}}{\frac{\mrm{d}}{\mrm{d} #1}}{\frac{\mrm{d}^#2}{\mrm{d} #1^#2}}}

\newcommand{\s}[1]{\langle #1 \rangle}

\newcommand{\mrm}[1]{{\mathrm #1}}

\def\l{\ell} % can clash with other definition
\def\tr{\mathrm{tr}}
\usepackage[normalem]{ulem}

\newcommand{\be}{\begin{equation}}
\newcommand{\ee}{\end{equation}}
\def\bea#1\eea{\begin{align}#1\end{align}}

  \def\th{\theta}

 \def\ab#1{\langle #1 \rangle}

\newcommand{\matrixel}[3]{\left< #1 \vphantom{#2#3} \right|
	#2 \left| #3 \vphantom{#1#2} \right>} % for Dirac matrix elements     

\usepackage{accents}
\newlength{\dhatheight}

\numberwithin{equation}{section}
\usepackage{tocloft}
\begin{document}

\vspace*{-.6in} \thispagestyle{empty}
\begin{flushright}
%CERN-TH-2017-124\\
%LPTENS \ldots
\end{flushright}
\vspace{1cm} {\Large
\begin{center}
\textbf{EFT  anomalous dimensions from the S-matrix  }
%\textbf{EFT  anomalous dimensions: \\[.2cm] two-loop leading logs from the tree-level S-matrix.}
\end{center}}
\vspace{1cm}
\begin{center}

{\bf  Joan Elias Mir\'o$^{a}$, James Ingoldby$^{a}$, Marc Riembau$^{b}$} \\[1cm]  {$^a$  ICTP,    Strada Costiera 11, 34135, Trieste, Italy  \\ $^b$    Univ. de Gen\`eve, Ernest-Ansermet 24, 1211 Gen\`eve, Switzerland \\ }
\vspace{1cm}%\today

\abstract{
We use the on-shell S-matrix and form factors to compute anomalous dimensions of higher dimension operators in the Standard Model Effective Field Theory. 
We find that in many instances, these computations are made simple by using the on-shell method. 
We first compute  contributions to anomalous dimensions of operators at dimension-six that arise
at one-loop. 
Then we calculate  two-loop anomalous dimensions  for which the corresponding one-loop contribution is absent, using this powerful method.
}

\vspace{3cm}%\today
\end{center}

 \vfill
 {
  \flushright
 \today 
}

\newpage

\setcounter{tocdepth}{2}

{
\tableofcontents
}

\section{Introduction}
\label{innn}

The discovery of the Higgs particle \cite{Chatrchyan:2012ufa,Aad:2012tfa}, Higgs coupling measurements, and a plethora of beyond the Standard Model (SM) searches suggest that there is an energy gap between the electroweak scale and  any new physics scale. Physics within this energy gap is appropriately described by the SM, together with a tower of operators encoding the deformations generated by the new dynamics.
Given such separation of scales, the  Renormalisation Group (RG) mixing of higher-dimension SM Effective Field Theory (EFT) operators can lead to important physical effects on precision observables.
The RG running of SM EFT operators down from a   hypothetical new physics scale to the EW scale is governed by the operator anomalous dimensions. 

The anomalous dimension of an operator can be computed using a number of different methods.
For instance,  the S-matrix elements, the effective action,  or form factors, are  subject to a Callan-Symanzik equation, a.k.a. renormalisation group equation (RGE), that depends on the anomalous dimensions. In this work we will compute the anomalous dimensions of SM dimension-six operators
 through the RGE's  satisfied by the form factors. 

The form factors  (FFs)  are defined by
\be
F_O( \vec{n}) \equiv \, {}_\text{out}\bra{\vec{n}} O(0) \ket{0}  \label{ffdef}\, ,
\ee
where $O(x)$ is a local and gauge invariant operator  and $ {}_\text{out}\bra{\vec{n}} $ is a multi-particle asymptotic state. These are the same  states  that are  used to  define the scattering $S$-matrix elements   
\be
S_{nm}\equiv \bra{\vec{n}}\hat S\ket{\vec{m}}=  \, _\text{out}  \bra{\vec{n}} \vec{m} \rangle_\text{in} \, . 
\ee
In perturbation theory, the calculation of the  FFs involves regularising  IR and UV divergences. 
For definiteness we will be using dimensional regularisation and the $\overline{\text{MS}}$ scheme so that the   FF computed  at a fixed order in perturbation theory  will depend on the 't Hooft scale $\mu$. 
 However, it  satisfies the Callan-Symanzik equation 
\be
%\frac{d}{d\mu}F_{O}( \vec{n};\mu) =  
\left( \mu \partial_\mu +\gamma -\gamma_\text{IR}+ \beta_{g}\partial_{g} \right) \, F_O( \vec{n};\mu)=0 \, ,   \label{CS1}
\ee
where  $\gamma$ is the anomalous dimension matrix,  $\gamma_\text{IR}$ is the IR anomalous dimension, and $\beta_g$ denotes collectively the beta function of the couplings in the theory. 
  $\gamma_\text{IR}$ is diagonal in the   space of kinematically independent operators and it is due to soft  and/or collinear emission of particles.

Next we  derive a relation between the form-factors and the S-matrix. We need three ingredients:
\begin{enumerate}
\item[1)]  Unitarity   $
F_O(\vec{n}) \equiv \, _\text{out}\bra{\vec{n}} O(0) \ket{0}  =    \sum_{\vec m} \, _\text{out}  \bra{\vec{n}} \vec{m} \rangle_\text{in} \, _\text{in}\bra{ \vec{m}} O(0) \ket{0}  $.
\item[2)]   The CPT theorem implies $ \, {}_\text{out}\bra{\vec{n}} O(0) \ket{0}  =   \bra{0} O^\dagger(0) \ket{\vec{n} }_\text{in} $. 
\item[3)]  Analyticity $F^*_O (s_{ij}-i \eps) = F_O(s_{ij}+i \eps) $. 
\end{enumerate}

Regarding $2)$, recall that CPT  is a symmetry transformation that is represented on the Hilbert space by an anti-linear and  anti-unitary operator. Thus   it relates in/out states when inserted in the inner product,   while acts on local operators   as  $\text{CPT}\,  O(x)\, \text{[CPT]}^{-1} = O^\dagger(-x)$. 
 Thus, point 2) follows from $\, {}_\text{out}\bra{\vec{n}} O(0) \ket{0}= \bra{0} \text{[CPT]}^{-1} \text{CPT}  \,  O^\dagger(0)  \text{[CPT]}^{-1} \text{CPT}  \ket{\vec{n} }_\text{in}$.

As for  $3)$, note that  due to Lorentz invariance $F_O(\vec n(s_{ij}+i\eps))\equiv F_O(s_{ij}+i\eps)$ must  depend on the Mandelstam invariants $s_{ij}=2p_i\cdot p_j$, for $i,j=1,\dots n$, which we collectively denote by $F_O(s_{ij}+i \eps)$, where $\eps>0$ is a small positive parameter arising from the Feynman $i \epsilon$-prescription.
The real analyticity relation in 3) is definitely true in perturbation theory because  complex conjugation of $F_O$ 
amounts to complex  conjugating the time-ordered propagators into anti-time-ordered propagators and thus $s_{ij}+i\eps\rightarrow s_{ij}-i\eps$. 
This  full counter-clockwise rotation 
in the $s_{ij}$ complex plane  can be generated by a complex dilatation of the momenta. Indeed, defining   the momentum space  dilatation generator
\be
D\equiv \sum_{\text{all particles}}  \,  p_i \frac{\partial}{\partial p_i} \, ,  \label{genD}
\ee
the relation in     $3)$  can be written as  $e^{-i \pi D}F^*_O (s_{ij}+i \eps) = F_O (s_{ij}+i \eps)$, which together with $1)$ and $2)$  implies
 \be
e^{-i \pi D} \, F^*_O ( \vec{n})  \, = \,    \sum_{\vec m} S_{nm}  \,  F^*_O ( \vec{m})    \, .  \label{msense0}
\ee 
This equation was  derived in  \cite{Caron-Huot:2016cwu}  using a  version of the optical theorem.
The derivation of \reef{msense0} that we present here   though  is  similar to  the proof of  the two-dimensional generalised Watson  theorem  \cite{Karowski:1978vz}.~\footnote{Watson's theorem \cite{Watson:1952ji,Watson:1954uc} is e.g. 
reviewed in \cite{Weinberg:1995mt}, formulated as  a relation between components of the S-matrix.  }
In two-dimensions complex conjugation in the rapidity plane $\th$ acts as  $
F^*_O(\th) = F_O(-\th)
$, where   $\th=|\th_{1}-\th_2|$ is the rapidity difference of two incoming particles of momenta $\{p_1,p_2\}$, and $(p_1+p_2)^2=4m^2 \cosh^2(\th_1-\th_2)/2$. 
 Then, by requiring only the extra assumption of a   factorised  two-dimensional S-matrix, equation \reef{msense0}  for $n=2$ reduces to the generalised Watson theorem $
F_O(\th)= F_O(-\th) \,  S(\th) $ \cite{Karowski:1978vz},    
where  $S(\th)$ denotes  the $2\rightarrow 2$ scattering element and we are omitting flavour indices;  this equation is exactly satisfied by many solvable two-dimensional QFTs.

 For the purposes of computing  UV anomalous dimensions,  one can consider all  particles as massless.  
  %In perturbation theory,  the FFs in  \reef{msense0} depend on the renormalisation scale $\mu$, and are subject to \reef{CS1}. 
  Next, following the observation of   \cite{Caron-Huot:2016cwu},  we relate     the Callan-Symanzik equation \reef{CS1} to  the  action of the momentum dilatation operator  
\be
DF_{O}\approx  - \mu \partial_\mu F_O^{(1)} = (\gamma-\gamma_\text{IR}+ \beta_{g}\partial_{g})^{(1)} F_O^{(0)}  \, . \label{cs1}
\ee  
Indeed, in dimensional regularisation there is a single renormalisation scale $\mu$ and we can trade derivatives of  $\sum_i p_i\partial_{p_i}$ by $-\mu \partial_\mu$.  
Thus, using  (\ref{msense0},\ref{cs1}) and expanding in powers of $D$ at first non-trivial order one gets~\cite{Caron-Huot:2016cwu} 
\be 
\bra{\vec{n}} O_j \ket{0}^{(0)}(\gamma_{ji}-\gamma_\text{IR}^i\delta_{ij})^{(1)} =-\frac{1}{\pi}\,  \bra{\vec{n}}  {\cal M} \otimes  O_i \ket{0}^{(0)} \, .  \label{chw1}
\ee
where as usual we have separated the S-matrix as $S=\mathds{1}+i {\cal M}$. 
The first factor on the left   $\bra{\vec{n}} O_j \ket{0}^{(0)}$ is a minimal form factor, namely the lowest order FF such that  does not vanish in the  free theory limit. The minimal FF is a polynomial in the kinematic variables: the momenta and polarisations of the particles.
The action of $\beta_g \partial_g$ can  be neglected because $\bra{\vec{n}} O_j \ket{0}^{(0)}$ does not depend on the  couplings of the theory. 
  On the right hand side,   `$\otimes$'  denotes an insertion of $\mathds{1}= \sum_{\vec m} | \vec{m} \rangle_\text{in}\,  \, _\text{in} \bra{\vec m} $ as in $2)$. 
 The ${\cal M}$-operator  and the  form factor on the right 
 are consistently  computed in perturbation theory so that it leads to a polynomial function like
 the l.h.s. minimal FF.  
 For instance, note that if the leading single $\log\mu$ in the FF of  the l.h.s. in \reef{msense0} arises at $\l$-loops, then applying  $-\mu\partial_\mu$ as  in \reef{cs1}  extracts the  anomalous dimension at $\l$-loop order.    
Thus `$(1)$' in  \reef{chw1} indicates the coefficient of the leading single $\log\mu$, which typically (but not necessarily) arises at one-loop. 
 From here on we will neglect the super-indices $(\cdots)^{(n)}$, $n=0,1$.

In section \ref{oneloopxcheck}   we will discuss the method
and compute  representative one-loop anomalous dimensions of  the SM dimension-six operators,  and compare the 
results with previous literature. 
In section  \ref{recap}  we  begin by summarising the pattern of vanishing entries in the matrix of one-loop anomalous dimensions. This allows us to identify those two-loop anomalous dimensions that are most interesting, namely those two-loop RG mixings that vanish at one-loop.  These can be efficiently obtained with the on-shell methodology and we compute   two-loop anomalous dimensions of this type  in section~\ref{easyQ}.  
Finally we summarise  our results and conclude in section~\ref{conc}.

\section{One-loop RGEs of dimension-six operators}
\label{oneloopxcheck}

In this section we apply  \reef{chw1} in a series of examples of one-loop  RGE mixing of the SM dimension-six operators. 
The one loop contributions to the right hand side of   \reef{chw1}  involve  a tree-level form factor and a tree-level S-matrix element, that are contracted  with a two-particle phase space integral.
Note that  the scattering matrix in the right hand side of  \reef{msense0} is not the fully connected S-matrix. 
Thus we should also consider contributions that involve a scattering matrix with disconnected pieces   \vspace{-.1cm}
\be
\matrixel{1, \dots , n}{O_j}{0}  \,  (\gamma_{ji}-\gamma_\text{IR}^i\delta_{ij}) =
  \begin{minipage}[h][1.3cm][t]{0.175\linewidth} \vspace{-.2cm} \diagTWO \end{minipage}=
 \sum_{k=2}^{n}  \begin{minipage}[h][1.07cm][t]{0.21\linewidth} \vspace{-.12cm} \diagTHREE \end{minipage}
  +\cdots   \label{chwgen} \, ,   \vspace{-.1cm}
\ee
 where   the lines to the right of the last  FF are the disconnected  particles of the  $S$-matrix.~\footnote{Indeed, from $
 \int \frac{d\text{LI}^\prime}{2\pi}  \,  \matrixel{1, \dots,  n}{{i{ \cal M}}}{ 1^\prime , 2^\prime, 3^\prime,  \dots, k^\prime }\matrixel{1^\prime , 2^\prime, 3^\prime,  \dots, k^\prime }{O_i}{0}$ % (where $d\text{LI}^\prime$ is the Lorentz invariant integral measure over  all the $p_i^\prime$ variables)
  pick  a connected $2\rightarrow k$ scattering element to reduce it to
 $ 
 \int \frac{d\text{LI}^\prime}{2\pi}  \,  \matrixel{1, \dots,  k}{{i {\cal M}}}{ 1^\prime , 2^\prime  }\matrixel{1^\prime , 2^\prime, k+1 ,  \dots, n}{O_i}{0}   
$.  The particles to the right of the last FF in \reef{chwgen} may involve tree-level interactions of the disconnected S-matrix, but we have absorbed those into the FF, such that both the S-matrix and the FF are connected. }
The dots in \reef{chwgen} denote the sum of  any possible permutation of the external particles,   and   higher loop contributions. 

For the purposes of computing the UV anomalous dimensions,  
one can consider all SM particles as massless, or equivalently  consider the SM in the symmetric phase of $SU(2)_L$. 
Then, since all the four-momenta involved in the problem \reef{chwgen} is null $p^2=0$, it is convenient to use  spinor helicity formalism \cite{Berends:1981rb,Berends:1981uq,Kleiss:1985yh}. 
An on-shell   massless spin $s$ particle is characterised by its helicity $h=\pm |s|$ and  a pair of spinors $\{\lambda_a, \tilde{\lambda}_{\dot{a}}\}$. The spinors
$\{\lambda_a, \tilde{\lambda}_{\dot{a}}\}$  transform in the representations $(1/2, 0)$ and $(0, 1/2)$ of the universal cover of the Lorentz group  $SL(2,\mathbb{C})$, respectively.
The basic invariant tensors are the antisymmetric Levi-Civita  $\epsilon^{ab}$,  $\epsilon^{\dot{a}\dot{b}}$,  and  $\sigma^\mu=( 1,\sigma^i)$ with $\sigma^i$ the Pauli matrices.
Given two particles with spinors $\{(\lambda_i)_a, (\tilde{\lambda_i})_{\dot{a}}\}$ where $i=1,2$, two simple Lorentz invariants are
\be
(\lambda_1)_a(\lambda_2)_b \eps^{ab}\equiv \braket{12}  \quad \text{and}  \quad (\tilde{\lambda}_1)_{\dot a}(\tilde{\lambda}_2)_{\dot b} \eps^{\dot a \dot b}\equiv [12] \, , \label{spin}
\ee
where we have introduced the commonly used angle and square bracket notation. Another basic invariant is
$
(p_i+p_j)^2 = \braket{ij}[ji]  \equiv s_{ij} \, , \label{spin2}
 $
where we are using mostly minus $\eta=(+{-}{-}{-})$ metric signature.  
Our convention  is  $p^\mu =\lambda^a (\sigma^\mu)_{a\dot a} \tilde{ \lambda}^{\dot a}$, and as usual we consider  $\lambda$ and $\tilde\lambda$ as independent complex spinors (i.e. two copies of $SL(2,\mathbb{C})$), however for real physical momenta (with positive energy) we should identify $\tilde\lambda=+\lambda^*$.
See for instance \cite{Henn:2014yza,Elvang:2015rqa} for further details on the spinor helicity formalism and on-shell scattering amplitudes techniques.

As usual, we factor out a  Dirac delta of  total momenta conservation from the interacting and connected  $i{\cal M}$ matrix
\be
\langle  12 \dots k  | i {\cal M}  | 1^\prime 2^\prime  \dots k^\prime \rangle =  (2\pi )^4 \delta^{(4)}\, 
  \begin{minipage}[h][.6cm][t]{0.02\linewidth} \vspace{-.12cm} \Big(  \end{minipage} 
  \text{\Large$\Sigma$}_{i=1}^k p_i  - \text{\Large$\Sigma$}_{i=1}^{k^\prime} p_i^\prime 
  \begin{minipage}[h][.6cm][t]{0.017\linewidth} \vspace{-.12cm} \Big) \end{minipage} 
 \,   i M( 1\dots  k ; 1 \dots k^\prime) \, ,
\ee
where $M$ is a function of the spinor variables $\{{\lambda_i}_a, {\tilde\lambda_i}_{\dot a}\}$ with $i\in \{1,\dots k\}$ and $\{{\lambda^\prime_i}_a, {\tilde\lambda^\prime_i}_{\dot a}\}$ with $i\in \{1,\dots k^\prime\}$ .

Starting with simplest things first, below we will show various one-loop computations that involve only the $2\rightarrow 2$ matrix elements in  first term of the r.h.s. of   \reef{chwgen}. 
Thus we need to perform the following type of integrals
\be
 \begin{minipage}[h][1.3cm][t]{0.175\linewidth} \vspace{-.12cm} \dQuatre \end{minipage} = 
  -\frac{1}{2\pi}\int d\text{LI}^\prime  \, 
  (2\pi )^4 \delta^{(4)}\,  ( p_1+p_2-p_1^\prime- p_2^\prime) \,  M( 12 ; 1^\prime 2^\prime) \, F_{O_i}(1^\prime 2^\prime 3  \dots  n)  \, ,  \label{tored}
 \ee 
where from here on we leave implicit the sum over all permutations of the external particles, and    $d\text{LI}^\prime$ is the Lorentz invariant integral measure over  all the $p_i^\prime$ variables, in this case  
$d\text{LI}^\prime = \prod_{i=1}^{2}\frac{d^3p'_i}{(2\pi)^3\, 2p_i'^0} $. 
The integral in \reef{tored} can be neatly performed sticking to the  spinor formalism  for the FF and the $M$ function. 
Since
$
p_1 + p_2 =  p_1^\prime+ p_2^\prime 
$ 
throughout the whole phase-space integration domain, it is useful to parametrise the  $\lambda_i^\prime$s in terms of the external particle's spinors by~\cite{Zwiebel:2011bx,Caron-Huot:2016cwu}
\be
\left( \begin{array}{c}   \lambda_1^\prime\\  \lambda_2^\prime  \end{array}\right) = 
%%%%%%
\left( \begin{array}{cc}  
 \cos \th & -\sin\th e^{i \phi}  \\
 \sin\th e^{-i \phi} & \cos \th  \\
\end{array} \right)
%%%%%%
\left( \begin{array}{c}   \lambda_1 \\  \lambda_2  \end{array}\right)  \, ,  \label{rot}
\ee 
and similarly for the $\tilde{\lambda}$'s spinors. 
Then, the two-particle phase-space  integral becomes an integral  over the rotation angles $\{\th,\phi\}$ only, and  \reef{tored} reduces to
\be
 F_{O_j}(12 \dots n) \,  (\gamma_{ji}-\gamma_\text{IR}^i\delta_{ij}) 
=  -\frac{1}{16\pi^2}\int_{0}^{2\pi}\frac{d\phi}{2\pi}\int_{0}^{\pi/2}2\sin\theta\cos\theta d\theta \,  M(1,2 ; 1^\prime 2^\prime)   \, F_{O_i}(1^\prime 2^\prime 3  \dots  n) \, .  \label{1loop}
 \ee
  For future reference we define the angular integration measure $d\Omega_2 \equiv \frac{d\phi}{2\pi} \, 2 \cos \theta \sin \th d\theta$.   See appendix \ref{psi} for  further details.

\subsection{Contributions at order $\lambda$}
\label{MFF}

In the following we compute the anomalous dimensions of the SM dimension-six operators at order $\lambda$  with 
the presented on-shell formalism.  
In  this case $\gamma_{IR}$ is absent since there are no collinear or soft divergences and thus using \reef{1loop} we can directly compute the UV anomalous dimensions. 
 
The tree-level $M$-matrix amplitude at $O(\lambda)$ due to $\delta{\cal L}= - \lambda |H|^4$ is given by
\be
M(1_i 2_j 3_k^* 4_l^*) =   - 2\lambda  \left( \delta_{i}^{\,   k}\delta_j^{ \,  l} + \delta_{j}^{\,   k}\delta_i^{ \,   l}   \right)  ,   \label{mat1}
\ee
where we have taken all the particles as outgoing and the indices run over the components of the   Higgs doublet $H$. 
By crossing outgoing particles into ingoing we generate the $2\rightarrow 2$ amplitudes  $M(1_i 2_j ; 1_k 2_l) $, $M(1_i 2_j^* ; 1_k 2_l^*)$, etc.

\subsubsection{$\partial^2 H^4$ operators}
We need to specify the  FFs that overlap with the scattering element in \reef{mat1}. 
Starting with operators involving four Higgses, there are two independent dimension-six operators
 $
O_{\|} =  |H |^2  |D^\mu H|^2$ and $O_{\bot} = |H^\dagger D_\mu  H|^2\, . 
 $ 
Their minimal FFs   with a four particle state are 
\bea
 F_{\|} (1_i 2_j 3_k^* 4^*_l) & =        \,  \delta_{i}^{\, l}\delta_{j}^{\, k}  \, \braket{14}[14]    + \,   \delta_{j}^{\, l}\delta_{i}^{\, k}   \, \braket{13}[13]   =
   \begin{minipage}[h][1.cm][t]{0.1\linewidth} \vspace{-.12cm} \dCinc{1_i}{2_j}{4_l^*}{3_k^*} \end{minipage}
+  \begin{minipage}[h][1.cm][t]{0.1\linewidth} \vspace{-.12cm} \dCinc{1_i}{2_j}{3_k^*}{4_l^*} \end{minipage}
\, ,  
\label{FF1}  \\%[.2cm]
 F_{ \bot}  (1_i 2_j 3_k^* 4^*_l)  & =        \, \delta_{i}^{\, l}\delta_{j}^{\, k}   \,   \braket{13}[13] +  \,   \delta_{j}^{\, l}\delta_{i}^{\, k}     \,  \braket{14}[14]    = 
 \begin{minipage}[h][1.cm][t]{0.1\linewidth} \vspace{-.12cm} \dSis{1_i}{4_l^*}{3_k^*}{2_j} \end{minipage}  +
 \begin{minipage}[h][1.cm][t]{0.1\linewidth} \vspace{-.12cm} \dSis{1_i}{3_k^*}{4_l^*}{2_j} \end{minipage}  \, ,       \label{FF2}
    %%%%%
\eea
   where we have used momentum conservation $\sum_{i=1}^4 p_i=0$.~\footnote{FFs  in momentum space, 
 $
F_O(p_1\dots p_n; q) \equiv  \int d^4x e^{iqx} F_O(p_1,\dots, p_n )$,   have the support of the four-momentum conservation delta function $\delta^{(4)}(q+\Sigma_i p_i)$. We will indistinguishably refer to $F_O(p_1\dots p_n; 0)$ by $F_O(p_1\dots p_n)$. } The two operators are in correspondence with the two possible ways to contract the momenta (solid line in (\ref{FF1},\ref{FF2}))  of two particles with two antiparticles that carry a flavour index (dotted line). 

Next we need to connect the FFs in  (\ref{FF1}-\ref{FF2}) with the matrix element \reef{mat1} in the various possible ways or channels \vspace{-.1cm}
\be
2 \begin{minipage}[h][1.5cm][t]{0.175\linewidth} \vspace{-.1cm} \LambUn{1_i}{2_j}{3_k}{4_l}{1^\prime_\alpha}{2^\prime_\beta} \end{minipage} + \
 4  \begin{minipage}[h][1.5cm][t]{0.175\linewidth} \vspace{-.1cm} \LambDos{1_i}{3_k^*}{4_l^*}{2_j}{3^{*\prime}_\beta}{1^\prime_\alpha}  \end{minipage}  + \
 4 \begin{minipage}[h][1.5cm][t]{0.175\linewidth} \vspace{-.1cm} \LambDos{1_i}{4_l^*}{3_k^*}{2_j}{4^{*\prime}_\beta}{1^\prime_\alpha}  \end{minipage}  
 \label{chans1}   \vspace{-.1cm}
 \ee
 where from here on we represent on-shell scalar particle lines with a dashed line, a crossed dot denotes an insertion of  the operator and a grey blob the ${\cal M}$-matrix.  
 Regarding the multiplicative factor in each channel, note that  a factor of two  is due to  crossing~\footnote{ I.e.  one gets the same result when exchanging  $\{ 1,2\} \leftrightarrow \{ 3,4\}$ ,  $\{ 1, 3\} \leftrightarrow \{ 2, 4\}$  and $\{ 1, 4\} \leftrightarrow \{ 2, 3\}$ in the first second and third channel respectively. },  while the last two channels get and extra factor of   two because    $|H_a, H_b^*\rangle$ and $|H_b, H_a^*\rangle$ are distinct states.

With this it is straightforward to compute the various channels.  Starting with the $F_{\bot}$ form factor, the first channel gives
 \be
-\frac{1}{16\pi^2} \int d\Omega_2 \, [   - 2 \lambda  ( \delta_{i}^{\,   \alpha}\delta_j^{ \,  \beta} + \delta_{j}^{\,   \alpha}\delta_i^{ \,   \beta}   ) ] \,   [     \, \delta_{\alpha}^{\, l}\delta_{\beta}^{\, k}   \,   \braket{1^\prime 3}[1^\prime 3] +  \,   \delta_{\beta}^{\, l}\delta_{\alpha}^{\, k}     \,  \braket{1^\prime 4}[1^\prime 4]  ]  \label{chan1} \, ,  \ee
where $d\Omega_2 $ is the angular integration in \reef{1loop}. 
The spinor $\langle 1^\prime |$ is rotated according to  \reef{rot}, which gives  $\ab{1^\prime i}=   \cos \th \ab{1  i}- \sin \th e^{i\phi} \ab{2i}$.
The $\phi$ integral sets to zero the terms with a phase $e^{i \phi}$, thus  
 $
\int d\Omega_2   \braket{1^\prime i}[1^\prime i] =  \int d\Omega_2 \,   ( \cos^2\th  \braket{1 i}[1 i] + \sin^2\th  \braket{2 i}[2 i] ) 
 $ 
for $i=3,4$.  After doing the $\th$ integral we get
\be
\reef{chan1}=   \frac{2 \lambda}{16\pi^2}   \left[
   \delta_{i}^{l }\delta_j^{ \, k} + \delta_{j}^{\,  l}\delta_i^{ \,  k}       \right]  \left[   \braket{1 3}[1 3] + \braket{1 4}[14]  \right]  \, .   \label{ts1}
\ee
The second  and third channels  in \reef{chans1} are done in the same way. The second one gives
\be
-\frac{1}{16\pi^2} \int d\Omega_2 \,  M(1_i 3_k^* ;  \, 1_\alpha^\prime 3^{*\prime}_\beta ) \, F_{\bot} ( 1_\a^\prime 2_j  3_\b^{*\prime}4_l^*) 
=  -  \frac{\lambda \braket{13}[13]}{16\pi^2} [  \delta_{i}^{\, k}\delta_{j}^{\, l} \, (N-1)  - 2  \delta_{i}^{\, l}\delta_{j}^{\, k}]  \, ,   \label{ts2}
\ee 
where $N=2$ is the dimension of the fundamental representation of the Higgs doublet. 
In this case we rotate $(|1^\prime \rangle , |3^\prime \rangle)= R(\th, \phi).  (|1\rangle , |3\rangle)$. 
Finally the third channel
\be
-\frac{1}{16 \pi^2 } \int d\Omega_2 \,  M(1_i 4_l^* ; \, 1_\alpha^\prime 4^{*\prime}_\beta ) \, F_{\bot} ( 1_\a^\prime 2_j  3_k 4_\beta^{*\prime}) 
= -  \frac{\lambda \braket{14}[14]}{16\pi^2} [  \delta_{i}^{\, l}\delta_{j}^{\, k} \, (N-1)   -  2  \delta_{i}^{\, k}\delta_{j}^{\, l}]  \, ,   \label{ts3}
\ee
can be obtained from the second channel  upon exchanging $3\leftrightarrow 4 $ and $k\leftrightarrow l $ on \reef{ts2}. After adding all the contributions, $ 2 \cdot \reef{ts1}+4 \cdot \reef{ts2}+4 \cdot \reef{ts3}$, we get
\be
    \frac{4\lambda  }{16 \pi^2 }   \,  [  \,   3 \,  F_{ \bot}  (1_i 2_j 3_k^* 4^*_l)+ (2-N)  \, F_{ \|}  (1_i 2_j 3_k^* 4^*_l)  \, ]  \, .    \label{fres}
\ee
 Thus for a SM Higgs doublet, we have that $N=2$ and the renormalisation  is diagonal.

Proceeding in a similar fashion for the $F_{\parallel}$ form factor we get the following result for the various channels: 
$t_1=     2\kappa  [ \delta_{i}^{\, l}\delta_{j}^{\, k}  + \delta_{i}^{\, k}\delta_{j }^{\, l} ]\braket{12}[12] $
, $t_2= - \kappa [      \delta_{i}^{\, k}\delta_{j}^{\, l} \, ( 2N+1 )   + \delta_{i}^{\, l}\delta_{j}^{\, k} ] \braket{13}[13]$ and 
 $t_3= - \kappa [     \delta_{j}^{\, k}\delta_{i}^{\, l} \, ( 2N+1 )   + \delta_{j}^{\, l}\delta_{i}^{\, k} ] \braket{14}[14]$, where $\kappa\equiv - 2\lambda (16\pi^2)^{-1}$. 
Adding the different channels, $ 2 t_1+4   t_2+4    t_3  $, gives
\be
  \frac{8\lambda}{16 \pi^2 }  \, (2N-1) \, F_{ \|}  (1_i 2_j 3_k^* 4^*_l)     \, .    \label{sres}
 \ee
From \reef{fres} and \reef{sres} we read the order $\lambda$ anomalous dimension matrix of the $\partial^2 H^4$-type SM operators 
$  \gamma_{c_{\bot}} = 12 \lambda  \gamma_{c_{\bot}} / 16 \pi^2  $ and 
$  \gamma_{c_{\|}} = 24 \lambda  \gamma_{c_{\|}} / 16 \pi^2  $,  
in agreement with  \cite{Grojean:2013kd,Elias-Miro:2013mua,Jenkins:2013zja}.~\footnote{
The renormalization mixing pattern in (\ref{fres},\ref{sres}) for $N=2$ can be understood using  custodial symmetry and spurion analysis \cite{Elias-Miro:2013mua}.   While $O_\bot$ transforms non-trivially, the quartic interactions and $O_\|$ are singlets under custodial transformation.   %For arbitrary integer $N> 2$ can be understood with a generalisation of  custodial symmetry.
 }

\subsubsection{The rest of order $\lambda$ corrections}

After having set the notation and basic steps, it is straightforward to compute the rest of order $\lambda$ corrections in the SM. We need to specify the minimal FFs that overlap with two Higgses
\bea
& F_{FF}(1_i 2_j^* 3_a^- 4_b^- )   =   -2 \delta_{i}^{\, j}   \delta_{AB} \,  \s{34}^2  \label{ffr1}  \, ,  \\[.2cm]
& F_6(1_a 2_b  3_c 4_d^* 5_e^* 6_f^* )     =   6% \left( 
 \ \delta_{a}^{\, d}  \delta_{b}^{\,  e} \delta_{c}^{\,  f}  
 +{\cal P}_{d,e,f}   \, ,   \\[.2cm]
& F_{y}( 1_i 2_j^* 3_k 4_l^- 5^-   ) =  ( \delta_{i}^{\, j} \delta_{k}^{\, l }  +  \delta_{k}^{\, j} \delta_{i}^{\, l }  )    \braket{45}  \label{OyMFF}   \, ,  \\[.2cm]
& F_{Q H} (1_i 2_j^* 3_k^{-} 4_l^+ )  = - T^{A}_{ji} T_{kl}^{A} \,  (   \braket{31}[14] - \braket{32}[24] )   \, ,  \label{ffr2}
\eea
where  ${\cal P}_X$ means  all the permutations of the previous term over the elements in  $X$.
The lower case and capital case indices  in (\ref{ffr1}-\ref{ffr2})  are fundamental and adjoint indices of $SU(2)$, respectively.
For operators containing fermions or vectors, one can generate FF with different helicity configurations. Throughout the text we will chose to do the calculations with  minimal FF that  have mostly negative helicity particles. Due to CPT,  if the calculation is done with the mostly positive helicity FF the results are the same. 

In  Table~\ref{tabbb} we provide the normalisation of the operators involved in these FFs. There we also provide the   rest of minimal FFs that we will use throughout out the paper. Note that when using on-shell methods, a natural base of dimensions-six operators to use is the so called Warsaw basis \cite{Grzadkowski:2010es}, because in such basis all operators that vanish on-shell have been traded by higher point contact interactions  \cite{Aoude:2019tzn,Durieux:2019eor,Falkowski:2019zdo,Durieux:2019siw,Ma:2019gtx}.

{ \renewcommand{\arraystretch}{1.45} \renewcommand\tabcolsep{6.7pt}
\begin{table}[h]%tbp]
\begin{center}
\begin{tabular}{L{1.cm} L{5.1cm}  L{3.4cm} L{2.7cm}  L{3.cm}  }  
\toprule  % \rowcolor{Gray2}
  & \bf{Operator } & \bf{MFF}    \\ \rowcolor{Gray1}
\midrule
      $ O_{3F}$     &   $\frac{f^{ABC} }{2\cdot 3!}F^{\mu}_{A \,\nu}F^{\nu}_{B\,\rho}\bar{F}^{\rho}_{C\,\mu} $ &     $F_3(1_A^-2_B^-3_C^-)$  &     $   \langle 12  \rangle   \langle 23 \rangle  \langle  31  \rangle$   &   $ i f^{ABC}/\sqrt{2}   $    \\  
      %%%
      $O_{FF}$           &   $ \frac{1}{2}H^\dagger H \,  F^A_{\mu\nu}\bar{F}^{ A \, \mu\nu}$ &     $ F_{FF}(1_i 2_j^* 3_A^- 4_B^- ) $   &    $   \langle 34  \rangle \langle 34  \rangle$     &       $- 2 \delta_{i}^{\, j}   \delta_{AB}$  \\   \rowcolor{Gray1}
      %%%%
       %  \hline
   $O_{qF}$      &   $\bar Q \sigma^{\mu\nu} T^A q H F^A_{\mu\nu}$ &      $ F_{qF}(1_i^- 2^- 3_k  4_A^- ) $      &   
   $ \langle 14  \rangle  \langle 42  \rangle$      &    $ 2 \sqrt{2} T^A_{ik} $   \\     
      %%%%%
         $O_{4F_1}$      &   $(\bar Q_i u  )\eps_{ij}( \bar Q_j d  ) $ &   $F_{4F_1} (1_i^-2^-3_j^-4^-)$    &     $ \langle 12  \rangle\langle 34 \rangle$     &    $\eps_{ij}  $
        \\  \rowcolor{Gray1}
   %%%%%%%%%
      $O_{y}$           &   $|H|^2 \bar Q q  H $ &   $F_{y}( 1_i 2_j^* 3_k 4_l^- 5^-   )$   &   $  \langle 45 \rangle $      &    $\delta  _{i}^{\, j}\delta  _{k}^{\, l}+\delta  _{k}^{\, j}\delta  _{i}^{\, l}$ \\   
      %%%%%%%
   $ O_{6}$     &   $|H|^6$ &   $F_6(1_a 2_b  3_c 4_d^* 5_e^* 6_f^* )$  &     $ 1 $    &    $ 6\, \delta_{a}^{\, d}  \delta_{b}^{\,  e} \delta_{c}^{\,  f}  
 +{\cal P}_{d,e,f} $  \\     \rowcolor{Gray1}
        %%%%%%
   $O_{4F_2}$      &   $(\bar  Q  T^A \gamma^\mu Q )(\bar Q T^A   \gamma_\mu Q ) $ &    $F_{4F_2} (1_i^-2_j^-3_k^+4_l^+)$     &    $    \langle 12  \rangle [   34]  $    &     $ 2T_{il}^{A} T_{jk}^{A }-2T_{jl}^{A} T_{ik}^{A } $  \\  
   %%%%%%
   $O_{Q H}$      &   $(\bar Q T^A  \gamma^{\mu} Q) \,  (i H^\dagger  T^A \lra{D}_\mu H)$  &     $F_{Q H} (1_i 2_j^* 3_k^{-} 4_l^+ ) $    &       $    \braket{31}[14]  $          &      $ - 2 \,  T_{ji}^{A} T_{kl}^{A }  $  
   \\  \rowcolor{Gray1}
   %%%%%   
$ O_{\bot}$     &   $ ( H^\dagger D_\mu H) ( D^\mu H)^\dagger  H$ &      $ F_{\bot} (1_i 2_j 3_k^* 4^*_l)$   &      $    \langle 13  \rangle [   13]  $      &    $\delta_{i}^{\, l }  \delta_{j}^{\, k } +{\cal P}_{1_i, 2_j}    $    \\  
%%%%%
$ O_{\|}$     &   $ |H|^2 ( D^\mu H)^\dagger (D_\mu H )$ &       $F_{\|} (1_i 2_j 3_k^* 4^*_l) $ &        $   \langle 14  \rangle [   14]$     &  $  \delta_{i}^{\, k }  \delta_{j}^{\, l } + {\cal P}_{1_i, 2_j} $ \\     
%%%%
   \bottomrule
\end{tabular}
\caption[Works]{ 
The Standard Model dimension-six operators and its minimal form factors (MFF) -- with the least number of particles.  The MFFs are given by the product of the  two rightmost  columns. 
In the right most column we give the $SU(2)_L$ and $SU(3)_c$ tensors that compound the MFF -- omitting the identity in colour space and flavour mixing matrices.
      $\bar F_{\a \b}=F_{\a \b} -i \eps_{ \a \b \mu \nu} F^{ \mu \nu}/2$,  and $T^A$ is either of  the generators of the SM gauge  groups.  
   We are not displaying  operators 
  whose MFF involves a straightforward modification of the tensor structure -- in the rightmost column -- of the operator shown. e.g.  like $H^\dagger \sigma^i H B_{\mu\nu}W^{i \, \mu\nu}$,   $i \tilde{H}^\dagger D_\mu H  (u \gamma^\mu d) $  and $|H|^2 \bar Q u \tilde H$, where  $\tilde{H}=i \sigma_2 H^*$.
     \label{tabbb} }
\end{center} \vspace{-.5cm}
\end{table}
}

Next we contract each  FF  in (\ref{ffr1}-\ref{ffr2}) with \reef{mat1} and  perform  the basic tensor multiplications and $d\Omega_2$ integrals as in \reef{1loop}.
Note that the calculation of the anomalous dimension of the FFs $F_{FF}$, $F_{6}$ and $F_y$ is rather simple   because the integration over the rotation angles is trivial and all what is left is to perform the contraction of the  $SU(2)_L$ tensors. 
Indeed, for $F_{FF}$ we get
\be
-\frac{2 \lambda}{16\pi^2 }   \left( \delta_{\a}^{\, j}\delta_i^{\, \b}+ \delta_{\a}^{\,  \b}\delta_{i}^{\, j}  \right)   \delta_{\b}^{\, \a} \delta_{ab} 2\s{12}^2  = \frac{2 \lambda}{16\pi^2 }       ( N+1)  \, F_{FF}(1_i 2_j^* 3_a^- 4_b^- ) 
\ee
The factor of $2$ arises because there are  two particle-antiparticle channels. 
Regarding the $O_6$ self-renorma\-li\-sation, there are  two  types of channels the particle-particle channel (or equivalently antiparticle-antiparticle) and the particle-antiparticle channel.  The are 6  particle-particle type channels, each of them   gives
 $ 
- M( 1_a 2_b ; 1_a^\prime 2_b^\prime)  F_6(1_{a}^\prime 2_{b^\prime}  3_c 4_d^* 5_e^* 6_f^*)   = 4 \lambda \,    F_6(1_{a} 2_b 3_c 4_d^* 5_e^* 6_f^*) $.
The second type,  the particle-antiparticle channel, 
gives 
 $
- M(1_a 4_d^* ;  1_a^\prime 4_d^{ \prime * })
 F_6(1_{a}^\prime 2_{b}  3_c 4_d^{\prime *} 5_e^* 6_f^* ) +{\cal P}_{d,e,f} = 14 \lambda   F_6(1_{a} 2_b 3_c 4_d^* 5_e^* 6_f^*)
 $  for a Higgs doublet. There are 6 types of contributions like this. Thus, all in all we  get 
 $
   6(4+14) \lambda  /(16 \pi^2)
$, 
at leading order in $\lambda$.
Regarding $F_y$, there are also two channels. The particle-particle    gives
$ 
- M( 1_i 3_k ; 1_a^\prime 3_b^\prime)  F_y(1_{a}^\prime 2_j^* 3^\prime_{b} 4_l^- 5^- ) = 4 \lambda F_y( 1_i 2_j^* 3_k 4_l^- 5^-   ) $. 
The second type,  the particle-antiparticle channel, 
gives 
$ 
- M(1_i 2_j^* ;  1_a^\prime 2_b^{ \prime * })F_y(1_{a}^\prime 2_b^{\prime *} 3_{b} 4_l^- 5^- )  + {\cal P}_{ik}= 2(N+3) \lambda F_y( 1_i 2_j^* 3_k 4_l^- 5^-   ) $; there are two such terms because $|1_a^\prime 2_b^{\prime *}\rangle$ and $|1_b^\prime 2_a^{\prime *}\rangle$ are distinct states. 
Adding all the channels  we get  $24 \lambda/(16 \pi^2)$ for the anomalous dimension  of  $F_y$.

As for the  calculation of $F_{QH}$,   there is only  a particle-antiparticle channel %\vspace{-.1cm}
\be
\begin{minipage}[h][1.3cm][t]{0.175\linewidth} \vspace{-.12cm} \LambDosAL{1_i}{2_j^*}{3_k^-}{4_l^+}{2^{*\prime}_\beta}{1^\prime_\alpha}  \end{minipage} \propto 
 \int d\Omega_2  \,   \braket{31^\prime}[1^\prime 4]   =   \braket{31}[14] + \braket{32}[2 4]     \, , \label{ttcite}   \vspace{-.1cm}
\ee
where the solid lines represent fermions. 
The former polynomial 
 vanishes by   momentum conservation $\sum_{i=1}^4| i\rangle [ i| = 0$.~\footnote{The corresponding diagram in an Effective Action calculation leads to $\partial |H|^2 \bar Q \gamma Q \stackrel{e.o.m.}{\longrightarrow} y O_y+ yO_y^*$, where $y$ is the Yukawa coupling and we have used the equations of motion. In the on-shell calculation, such order $\lambda y$ contribution comes from using $\langle HH^\dagger H \bar Q Q |$ as an outgoing state in \reef{1loop}. }
This is of course unsurprising since the mixing is only sensitive to the  current $J^A_\mu=i   H^\dagger  T^A \partial_\mu H-  i  \partial_\mu H^\dagger  T^A H $ in the operator  $O_{QH}=J^A_{ \mu}  (\bar Q \gamma^\mu  T^A Q ) $. 
The current $J^A_\mu$ is conserved (up to terms proportional to the  $SU(2)_L$ coupling $g$), hence it obeys a Ward identity and thus does not acquire an anomalous dimension proportional to $\lambda[1+ O(g)]$.

All in all, 
we recover the order  $\lambda$ anomalous dimension matrix of the  
dimensions-six operators
\be 
\left( \begin{array}{l}  \gamma_{FF} \\   \gamma_{\bot} \\    \gamma_{\|}  \\   \gamma_6 \\   \gamma_y \\   \gamma_{\psi H } \end{array}\right)= 
%%%%%%
\frac{4 \lambda }{16 \pi^2}
\left( \begin{array}{cccccc}  
3 &  &  & & &   \\
 & 3 &   & & &   \\
  &  & 6 &  & &  \\
  &  &  &  27 & &  \\
  &  &  &  &  6 &  \\
  &  &  &  &   &  0
\end{array} \right) 
%%%%%%
\left( \begin{array}{l}  c_{FF} \\  c_{\bot} \\   c_{\|}  \\  c_6 \\  c_y \\  c_{\psi H } \end{array}\right) \, , \label{agrr}
\ee 
where  we only show the non-vanishing matrix elements. Eq.~\reef{agrr} is in agreement with \cite{Grojean:2013kd,Elias-Miro:2013mua,Jenkins:2013zja}.

\subsection{Contributions from gauge interactions}

Now we consider contributions to anomalous dimensions that are proportional to a gauge coupling squared. These contributions arise when an $O(g^2)$ ${\cal M}$-matrix element is used in \reef{1loop}. In this case however, ${\cal M}$-matrix elements can have collinear and soft divergences, which generate a non-vanishing contribution to $\gamma_\text{IR}$. We will therefore describe how the formalism must be used when IR divergences are present, first calculating the anomalous dimension of a conserved Noether current as a simple example. Then, we will calculate gauge contributions to the anomalous dimensions of $O_\|$ and $O_\bot$.

\subsubsection{IR divergences}

In a %any 
gauge theory, the IR anomalous dimension  
has the following expression at the one-loop level:
\be
\gamma_\text{IR}(s_{ij} ; \mu) =    \sum_{i < j } T^A_i \cdot T^A_j  \,  \gamma_\text{cusp} \,   \log  \frac{\mu^2 }{-s_{ij}} + \sum_{i}  \gamma^\text{coll}_{i}  \, , 
\label{alll}
\ee 
see e.g. refs.~\cite{Becher:2009cu,Becher:2014oda}.  In \reef{alll}  the sum runs over ordered pairs of external legs, $T^A_i$ is a generator of the gauge group, and $\mu$ is the t'Hooft scale of dimensional regularisation. This form for $\gamma_\text{IR}$ has also been confirmed to hold in a number of examples at two and three loops in QCD and ${\mathcal N}=4$ SYM (see \cite{Becher:2014oda} and references therein), and has been conjectured to hold at all subsequent loop orders. For our current purposes it suffices that \reef{alll} holds at one loop for either of the SM gauge groups with collinear dimensions given by
\begin{equation}
	\gamma^\text{coll}_g = -b_0\frac{g^2}{16\pi^2},\qquad \gamma^\text{coll}_q = -3C_2\frac{g^2}{16\pi^2},\qquad \gamma^\text{coll}_{H} =   -4C_2\frac{g^2}{16\pi^2},\label{colldims}
\end{equation} 
where $b_0$ is the one-loop beta function coefficient, $C_2 \mathds{1}=\sum_A T^{A}T^A$, and at one loop $\gamma_\text{cusp}=g^2/(4\pi^2)$. The collinear dimensions $\gamma^\text{coll}_{g/q}$ are shown in \cite{Becher:2009cu} and $\gamma^\text{coll}_H$ can be extracted from \cite{Chiu:2009mg}.
Using \reef{alll} in \reef{tored} it can be shown that~\cite{Caron-Huot:2016cwu}
\be
\gamma_{ji} F_{O_j} 
=  -\frac{1}{16\pi^2}\int d\Omega_2 \,  \left( M(12;1^\prime 2^\prime)F_{O_i}(1^\prime 2^\prime 3  \dots  n)  +\frac{2g^2 T_{1}^A T^A_{2}F_{O_i}(123\dots  n)}{\sin^2\th \cos^2\th }\right)  \,+ F_{O_i} \sum_k \gamma_k^\text{coll} \, ,  \label{genir}  
\ee
where as in \reef{tored} we leave implicit the sum over permutations of ordered pairs of the $n$ external particles.

The FF of a Noether current that corresponds to an ungauged global symmetry has a vanishing UV anomalous dimension. Next, we will calculate the anomalous dimension of such a FF and confirm that it is zero to provide a good consistency check on \reef{genir}. We have an ungauged global symmetry in the SM Higgs sector, associated with $SU(2)$ rotations of the Higgs doublet once the $SU(2)_L$ gauge coupling has been set to zero. The corresponding Noether current is
$J^A_\mu=   i (H^\dagger  T^A \partial_\mu H-   \partial_\mu H^\dagger  T^A H  )$, and its associated  MFF is 
\be
F_{J^A}(1_i 2_j^*) =  - T^A_{ij}   \,  \big(\lambda^\a_1\tilde{\lambda}^{\dot{\a}}_1-\lambda^\a_2\tilde{\lambda}^{\dot{\a}}_2 \big) \label{cFF}  \, , 
\ee
where $T^A = \sigma^A/2$.
We will consider the contribution to the UV anomalous dimension of this FF coming from the SM Hypercharge interaction. To compute this contribution to $O(g^{\prime 2})$, we need the tree--level $2\rightarrow2$  scattering matrix element for Higgs scalars due to the Hypercharge gauge interaction:
\begin{equation}
	M(1_i 2_j 3^*_k 4_l^*)= g^{\prime 2}Y_H^2  \, \delta_{i}^{\, k}\delta_{j}^{\, l}\left(1-2\frac{\langle 12\rangle\langle 34\rangle}{\langle 13\rangle\langle 24\rangle}\right)+    g^{\prime 2}Y_H^2 \, \delta_{i}^{\, l}\delta_{j}^{\, k}\left(1-2\frac{\langle 12\rangle\langle 34\rangle}{\langle 23\rangle\langle 41\rangle}\right) \, . \label{scalqedM}
\end{equation} 
Evaluating \reef{genir} for the non-abelian current in \reef{cFF}, we have
\begin{multline}
	\gamma_{JJ}F_{J^A}( 1_i2^*_j )= -\frac{1}{16\pi^2}    \int d\Omega_2 \,   \bigg(  M (1_i2^*_j ; 1'_\alpha 2^{*'}_\beta)F_{J^A} (1'_\alpha 2^{*'}_\beta ) + M (1_i2^*_j ; 1^{*'}_\alpha 2^{'}_\beta)F_{J^A} (1^{*'}_\alpha 2^{'}_\beta ) \\-\frac{g^{\prime 2} F_{J^A}( 1_i2^*_j )}{2\sin^2\th \cos^2\th }\bigg)      + F_{J^A} ( 1_i2^*_j )\sum_k \gamma_k^\text{coll}, 
	\label{xir}
\end{multline}
where we have used $T^A_1T^A_2=-1/4$ for the hypercharge generators  $Y_H=1/2$ acting on a Higgs scalar and its antiparticle. 
The integral in \reef{xir} is evaluated as usual. After performing the angular rotations and basic integrations, 
we find that it equals $2g^{\prime 2}F_{J^A}( 1_i2^*_j )/16\pi^2$. 
Next, using the result for the collinear dimension of scalars from \reef{colldims}, and that $C_2=1/4$, we find that the final term in Eq.~(\ref{xir}) equals $-2g^{\prime 2}F_{J^A}( 1_i2^*_j )/16\pi^2$.
We have therefore demonstrated that the required result $\gamma_{JJ}=0$, can be recovered using this formalism.

In passing, we note that if the same exercise is instead carried out for the Noether current associated with the gauged $U(1)_Y$ symmetry, we get a non-vanishing anomalous dimension. This is because renormalisation of the naive Noether current occurs when the corresponding abelian symmetry is gauged. However, an improved current can be defined in this case that is both conserved and non-renormalised \cite{Collins:2005nj}.

Having done some simple checks of \reef{genir}, we now compute instances of gauge contributions to  the SM dimension-six anomalous dimensions.

\subsubsection{Gauge mixing}

Coming back to the operators $O_\|$ and $O_\bot$, let us discuss the mixing due to the  $U(1)_Y$-Hypercharge. We will use  (\ref{genir}) to calculate the matrix of anomalous dimensions, which is needed as there is both operator mixing and a non-vanishing IR anomalous dimension for this interaction and choice of FFs. In this formula, we will plug in the matrix element for scalar scattering due to the Hypercharge interaction, given by (\ref{scalqedM}).

Begin by first taking $F_\bot$ in the integral. Just like the corresponding $O(\lambda)$ calculation in section~\ref{MFF}, there will be three independent channels that contribute to the integral, corresponding to different pairs of scalars crossing the cut. The integral for the first channel gives
\bea
	-\frac{1}{16\pi^2} &\int  d\Omega_2 \left(M(1_i2_j;1'_\alpha 2'_\beta)F_\bot(1'_\alpha 2'_\beta 3^*_k4^*_l)+\frac{g^{\prime 2}F_\bot(1_i2_j3^*_k4^*_l)}{2\sin^2\theta\cos^2\theta}\right)  \nonumber \\ &= -  \frac{g^{\prime 2}}{64\pi^2}\delta_i^l\delta_j^k\left\{5\braket{13}[31]-3\braket{14}[41]\right\}-\frac{g^{\prime 2}}{64\pi^2}\delta_j^l\delta_i^k\left\{5\braket{14}[41]-3\braket{13}[31]\right\},\label{Eq:gprime12}
 \eea
where the Higgs Hypercharge value $Y_H=1/2$ has been inputted. Individually, the two terms appearing in the integral in Eq.~(\ref{Eq:gprime12}) are divergent, but their sum is finite.
The integral for the second channel is
\bea
	-\frac{1}{16\pi^2} & \int  d\Omega_2 \left(M(1_i3^*_k;1'_\alpha 3^{*\prime}_\beta)F_\bot(1'_\alpha 2_j 3^{*\prime}_\beta 4^*_l)+M(1_i3^*_k;1^{*\prime}_\alpha 3^{\prime}_\beta)F_\bot(1^{*\prime}_\alpha 2_j 3^{\prime}_\beta 4^*_l)-\frac{g^{\prime 2}F_\bot(1_i2_j3^*_k4^*_l)}{2\sin^2\theta\cos^2\theta}\right)  \nonumber\\
&	=\frac{2g^{\prime 2}}{64\pi^2}\delta_i^l\delta_j^k\braket{13}[31]+\frac{g^{\prime 2}}{64\pi^2}\delta_j^l\delta_i^k\left\{\left(\frac{N}{3}+3\right)\braket{13}[31]+\left(\frac{2N}{3}+8\right)\braket{14}[41]\right\}.
 \eea
The integral for the third channel can be found from the second by making the substitutions $3\leftrightarrow 4$ and $k\leftrightarrow l$.
Finally, multiplying each of these three channels  by two (to include the crossing symmetric contributions) and adding them gives the total value for the integral in (\ref{genir}). Using the equations for the FFs in  (\ref{FF1}) and (\ref{FF2}), we can express this total as
\be
\frac{g^{\prime 2}}{16\pi^2}\left(\frac{N}{3}+\frac{5}{2}\right)F_\bot(1_i2_j3^*_k4^*_l)+\frac{g^{\prime 2}}{16\pi^2}\left(\frac{N}{6}+3\right)F_\|(1_i2_j3^*_k4^*_l) \, . 
\ee
Using (\ref{colldims}), we find that the collinear contribution is $\sum \gamma^\text{coll}=-4g^{\prime 2}/16\pi^2$. Then by matching coefficients of $F_\bot$ and $F_\|$ in equation (\ref{genir}), we can extract order $g^{\prime}$ values for two elements of the anomalous dimension matrix -- $\gamma_\bot = g^{\prime 2}(N/3-3/2)c_\bot/16\pi^2$ and $\gamma_\|=g^{\prime 2}(N/6+3)c_\bot/16\pi^2$.

By repeating this procedure except inputting $F_\|$ to the integral in (\ref{genir}), the remaining two elements of the anomalous dimension matrix can be found. After combining the three channels in the same way as before, the integral term is
\be
\frac{g^{\prime 2}}{16\pi^2}\left(\frac{10}{3}F_\bot(1_i2_j3^*_k4^*_l)+\frac{8}{3}F_\|(1_i2_j3^*_k4^*_l)\right),
\ee 
independently of $N$. After adding the collinear contribution and matching coefficients of  $F_\bot$ and $F_\|$ in Eq.~(\ref{genir}), we can find the remaining two elements of the $O_\|$ and $O_\bot$ anomalous dimension matrix at order   $g^{\prime 2}$. After setting $N=2$ for the Higgs doublet, the final result is  
\be
\begingroup
\renewcommand*{\arraystretch}{1.5}
\begin{pmatrix}
	\gamma_\bot\\
	\gamma_\|
\end{pmatrix}
=\frac{g^{\prime 2}}{16\pi^2}\begin{pmatrix}
	-\frac{5}{6} && \frac{10}{3} \\
	\frac{10}{3} && -\frac{4}{3}
\end{pmatrix}
\begin{pmatrix}
	c_\bot \\
	c_\|
\end{pmatrix},
\endgroup
\ee 
in agreement with previous literature \cite{Elias-Miro:2013mua,Alonso:2013hga}.~\footnote{The calculation of \cite{Elias-Miro:2013mua} was done in a basis where    $O_B\equiv (H^\dagger D_\nu H) \partial_\mu B^{\mu\nu}$ is kept as an independent operator. After rotating away such operator and rotating the operators $\{O_H,O_T\}$ (there defined) into  $\{ O_\| , O_\bot \}$ full agreement is found. An early calculation of the RG mixing from $O_H$ to $O_T$ can be found in \cite{Barbieri:2007bh}.} 
By similar means one can compute the mixing due to the $SU(2)_L$ gauge interactions.  However, such order $g^2$ contributions are diagonal because,  due to  custodial symmetry, the RG mixing between $O_\|$ and $O_\bot$ must be proportional to the Hypercharge coupling $g^{\prime 2}$  \cite{Elias-Miro:2013mua}.

\subsection{Non-minimal form factors}
\label{nonMFF}

In the previous sections we have only  encountered minimal form factors, i.e. FF that do not vanish at zero order in the SM couplings. 
This is because  we focused on transitions between operators that have the same number of external fields. Instead, if the transition between operators increases the number of particles, we will potentially have contributions from a $2\to 2$ S-matrix element acting on a non-minimal form factor.%,  a.k.a. a tree-level FF. 

Next we  compute an illustrative example consisting on the dipole renormalisation by the $F^3$ operator. For concreteness we focus on the $F^3$ and dipole involving the gluon field strength. 
The MFFs that  we  need are~\footnote{Where we used the convention $\eps^-_{a\dot a} = -\sqrt{2}  \lambda_a \tilde{\mu}_{\dot a} /[\tilde \lambda,\tilde \mu] $.}  
\be
 F_{3G}(1_A^-2_B^-3_C^-) =  \frac{i}{\sqrt{2}} f^{ABC} \,\ab{12}\ab{23}\ab{31}  \ ,   \quad F_{qG}(1^-_i 2^-_j 3\, 4^-_A) = 2\sqrt{2}\,T^A_{ij} \,\ab{14}\ab{42} \label{FF3G}  \, ,
\ee
where in $F_{qG}$ we only show the color index, because the $SU(2)$ indices are diagonal throughout the calculation. 
There are three different contributions, 
\be
\begin{minipage}[h][1.07cm][t]{0.235\linewidth} \vspace{-.12cm} \dNMFFun \end{minipage} +  
 \begin{minipage}[h][1.07cm][t]{0.23\linewidth} \vspace{-.12cm} \dNMFFdos \end{minipage}   +
    \begin{minipage}[h][1.07cm][t]{0.23\linewidth} \vspace{-.12cm} \dNMFFtres \end{minipage}    \, . 
	\label{f3todipdiags}
\ee
The first diagram in   \reef{f3todipdiags}  involves a 5-point scattering amplitude.
The amplitude between two fermions, a scalar and two gluons with all-plus helicity vanishes, $M(\psi^+ \psi^+ \phi\, g^+ g^+)=0$. Therefore the only contribution is from the amplitude with negative helicity fermions $M(\psi^-\psi^- \phi\, g^+g^+)$, which is a potential contribution to the dipole  $F_{qG}(1^-_i 2^-_j 3\, 4^-_A)$ in \reef{FF3G}. 
The 5-point amplitude is
\be
M(1^-_i 2^+_A 3^+_B 4^-_j 5\,)\,=\,  -2 y g_s^2 (T^AT^B)_{ij}\,\frac{\ab{14}^2}{\ab{12}\ab{23}\ab{34}}\,
-2 y g_s^2 (T^BT^A)_{ij}\,\frac{\ab{14}^2}{\ab{13}\ab{32}\ab{24}} \, , 
\label{ampffsgg}
\ee
We need to perform the convolution of this amplitude with  $F_{3G}$ in   \reef{FF3G}, 
\be
-\frac{1}{16\pi^2}\,\int \text{d}\Omega_2  \,  
M( 1^-_i 3^-_j 4 ; x^-_\mathsf{X}y^-_\mathsf{Y}  ) \, 
 F_{3G}(x^-_\mathsf{X}y^-_\mathsf{Y} 2^-_A)  = -\frac{yg_s^2}{16\pi^2}\frac{ N_c T^A_{ij}}{\sqrt{2}}2\ab{13}^2\int \text{d}\Omega_2  \frac{\ab{y2}\ab{2x}}{\ab{1x}\ab{y3}}  \label{ttcc}
\ee
where we used $-if^{ABC}T^AT^B=\frac{1}{2}N_c T^C$,  and a factor of two arises because   both color orderings  in \reef{ampffsgg} give  the same integral since it is invariant under $x\leftrightarrow y$. Now we need to rotate the internal spinors in terms of the three external ones. 
We rotate $|x\rangle$ and $|y\rangle$ in terms of spinors $|a\rangle$ and $|b\rangle$ such that they are on-shell and $p_a+p_b=p_1+p_3+p_4$, a simple choice is~\cite{Caron-Huot:2016cwu}
\be
|a\rangle = |1\rangle \sqrt{\frac{s_{134}}{s_{13}+s_{14}}}\,,
\hspace{1cm} 
|b\rangle = (|3\rangle [13]+|4\rangle [14]) \frac{1}{\sqrt{s_{13}+s_{14}}}\,
\label{lambdaalambdab}
\ee
Next we rotate $(|x\rangle, |y \rangle)=R(\th,\phi).(|a\rangle, |b \rangle)$ like in \reef{rot} and we are lead to 
\be
\int_0^{\pi/2}2c_\theta s_\theta\text{d}\theta\int\frac{\text{d}\phi}{2\pi}\,
\frac{(\ab{b2}c_\theta+s_\theta e^{-i\phi}\ab{a2})(\ab{2a}c_\theta-s_\theta e^{i\phi}\ab{2b})}{-s_\theta e^{i\phi}\ab{1b}(\ab{b3}c_\theta+s_\theta e^{-i\phi}\ab{a3})} \, .   \label{citar1}
\ee
A convenient way to perform the $\phi$ and $\theta$ integrals is to change variables to $z\equiv e^{i\phi}$ and $t\equiv \tan\theta$. In this way, the denominator of the integrand is a product of terms like $(z-\ab{ij}/\ab{ik} t)$ so the $z$ integral picks the residues inside the unit circle and the integral on $t$ scans the position of those residues. The integral in \reef{citar1} is
\be -2\frac{\ab{2b}\ab{2a}}{\ab{1b}\ab{3b}}\int_0^\infty \frac{\text{d}t}{(1+t^2)^2}\oint_{|z|=1}\frac{\text{d}z}{2\pi i}\frac{(z+t\, r_2)(1-z t/r_2)}{z^2(z+t \,r_3)}  \label{cctt2}\, , 
\ee
where we have defined $r_i = \ab{ia}/\ab{ib}$. After performing the integrals, we use the expressions in \reef{lambdaalambdab}, and after a lengthy manipulation of the result using Schouten identities $[\xi i][jk]+[\xi j][ki]+[\xi k][ij]=0$  we get
\be
 \reef{cctt2}
=
\frac{\ab{12}\ab{23}}{\ab{13}^2}\left(  2 -\log\frac{(s_{13}+s_{14})(s_{13}+s_{34})}{s_{14}s_{34}} \right) + \frac{\ab{12}^2\ab{34}}{\ab{14}\ab{13}^2}\frac{s_{14}}{s_{13}+s_{14}}
+ \frac{\ab{14}\ab{23}^2}{\ab{34}\ab{13}^2}\frac{s_{34}}{s_{13}+s_{34}} \, .
\label{eq:minfff3todipole}
\ee

Next we  compute  the right diagram in  \reef{f3todipdiags}, which involves the following four-particle amplitude and   non-minimal FF  
\be
M(1^-_i 2^+_A 3^-_j 4\,) \,=\,
yg_s \sqrt{2} T^A_{ij}\, \frac{\ab{13}^2}{\ab{12}\ab{23}}\,
\hspace{0.5cm}\text{and}\hspace{0.5cm}
F_{3G}(1^-_i 2^-_A 3^-_B 4^+_j)  = -i g_s f^{ABC} T^C_{ij}\, \frac{\ab{12}\ab{23}\ab{31}}{\ab{34}} \, . 
\ee
Convoluting this scattering matrix element with the FF we get
\bea
-\frac{2}{16\pi^2}\int \text{d}\Omega_2 M( 1^-_i  4 ;x^-_\mathsf{X}y^+_k ) \, F_{3G} (   2^-_A x^-_\mathsf{X} 3^-_jy^+_k )  =  
-\frac{yg_s^2}{16\pi^2}\frac{ N_c T^A_{ij}}{\sqrt{2}}2\,\frac{\ab{23}}{\ab{14}}\int \text{d}\Omega_2  \frac{\ab{1y}^2\ab{2x}\ab{3x}}{\ab{1x}\ab{3y}}. \label{ttccc2}
\eea
The factor of two comes because  there are two  channels  corresponding to propagating either  $|x^-_\mathsf{X}y^+_k\rangle$  or  $|y^+_k x^-_\mathsf{X} \rangle$,  that give the same value.  
Next we rotate $(|x\rangle, |y \rangle)=R(\th,\phi).(|1\rangle, |4 \rangle)$ and perform again  the change of variables into $z\equiv e^{i\phi}$ and $t\equiv \tan\theta$.
We are left with the following integral 
\be
2\int_0^\infty \frac{\text{d}t}{(1+t^2)^3}\oint_{|z|=1}\frac{\text{d}z}{2\pi i}\frac{(1+z\,t\,\frac{\ab{24}}{\ab{12}})(\frac{\ab{13}}{\ab{34}}+t\,z)}{z(t\frac{\ab{13}}{\ab{34}}-z)}  = \left( -\frac{3}{2}+\log\frac{s_{13}+s_{34}}{s_{34}} \,-\,\frac{\ab{14}\ab{23}}{\ab{12}\ab{34}} \frac{s_{34}}{s_{13}+s_{34}} \right)  \, .
\label{nonminfff3todipole} 
 \ee
 The final result for this contribution is  $-\frac{yg_s^2}{16\pi^2}\frac{ N_c T^A_{ij}}{\sqrt{2}}  \,  2\ab{12}\ab{23} \times \reef{nonminfff3todipole}$.

The third   contribution in \reef{f3todipdiags} comes from exchanging $1$ and $3$ in  the second contribution.  
Finally  adding  the three  contributions, we see that  the $\log$'s and rational functions cancel and we are left with the   contact interaction
 \be
\frac{yg_s^2}{16\pi^2} \frac{ N_c }{2}   \,   T^A_{ij} 2 \sqrt{2}\ab{12}\ab{23}\, , 
 \ee
which we easily recognise as the dipole MFF in \reef{FF3G}. Thus   the
anomalous dimension is
\be
\gamma_{qG\leftarrow 3G} \, =\, \frac{g_s^2 y}{16\pi^2} \,\frac{N_c}{2} \, ,  \label{serr}
\ee
in agreement with \cite{Braaten:1990gq}. 

In passing we note that if we do the calculation with the  positive helicity gluons $ F_{3G}(1_A^+2_B^+3_C^+)$, i.e. the CPT related FF, we get the complex conjugate dipole. 
The Wilson coefficient   in  $c_{3G}O_{3G}$ is taken as complex and thus we generate the operator $O_{qG}$  with a complex coefficient, namely the CP even and CP odd dipole.

\section{Structure of the anomalous dimension matrix}
\label{recap}

The on-shell formalism that we have been using suggests classifying the  operators -- or minimal FF --  in five classes, 
\be
\ab{ \ \cdot \ }^3 \, ,  \quad  \ab{ \ \cdot \ }^2 \, ,  \quad  \ab{ \ \cdot \ } \, ,  \quad  1 \quad \text{and}    \quad   \ab{ \ \cdot \ }[ \ \cdot \ ] 
\ee
 depending on the number of $\lambda$ and $\widetilde{\lambda}$'s. For instance $F_{3G}(1_A^-2_B^-3_C^-)\propto \ab{12 } \ab{23}\ab{31} \sim \ab{ \ \cdot \ }^3$ and therefore is of the first class, see table \ref{tabbb} for the rest of FFs. 
The total number of spinors and dimensional analysis  determines the number of particles involved in each FF.~\footnote{This coincides with the classifications in  \cite{Cheung:2015aba,Henning:2019enq}.}

The on-shell  formalism makes very transparent  the  pattern of the anomalous dimensions matrix, namely the non-vanishing entries of the matrix. 
When mixing  from $O_i$ to $O_j$, the phase-space integral will remove two legs from the form factor $F_{O_i}$ 
while the $S_{m\leftarrow 2}$ scattering matrix element generates $m\geq 2$ legs. Thus  $O_j$ has, the same number of external fields or more, but never less than $O_i$. Therefore operators can only renormalise,  at one-loop, operators  with equal or more fields. This is an instance of the more general theorem in \cite{Bern:2019wie}.

The RG mixing between the two four-particle classes --  $\ab{ \ \cdot \ }^2$ and  $ \ab{ \ \cdot \ }[ \ \cdot \ ] $ -- 
requires the 2 to 2  S-matrix element to  have a total net helicity $\sum_i h_i\neq 0$ with all particles outgoing.
In the SM there is a unique four-point tree-level amplitude that has net helicity, namely the four-fermion amplitude proportional to the up and down Yukawas $y_u y_d$  \cite{Cheung:2015aba,Azatov:2016sqh}. 
Therefore the only allowed mixing between operators from class $\ab{ \ \cdot \ }^2$ and   $ \ab{ \ \cdot \ }[ \ \cdot \ ] $ is between the four-fermion operators $\psi^4$ and $\psi^2 \bar\psi^2$. This effect was previously explained using  SUSY as a spurious symmetry, since the up and down Yukawa interactions are not simultaneously holomorphic  \cite{Alonso:2014rga,Elias-Miro:2014eia}. 
Finally, helicity selection rules \cite{Cheung:2015aba}, or accidental supersymmetry \cite{Elias-Miro:2014eia},   also exclude one-loop mixing from operators in  $\ab{ \ \cdot \ }^3$, into  $\ab{ \ \cdot \ }[ \ \cdot \ ]$.~\footnote{Helicity selection rules also imply that the corresponding one-loop diagram vanishes up to a rational function  of the kinematical variables; in many cases the rational factor  vanishes as well \cite{Craig:2019wmo}, see also below.}

 \begin{figure}
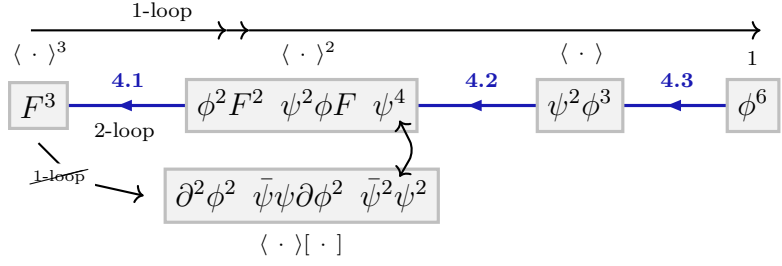

	\centering
 \begin{minipage}[h][3.55cm][t]{0.52\linewidth} \vspace{-.12cm} \dMaster  \end{minipage}  
	\caption{One-loop RG mixing can only occur from operators renormalising operators in boxes to the right or in the same box, with the only exception of the four-fermion of type $\ab{ \ \cdot \ }^2$  and type  $\ab{ \ \cdot \ }[ \ \cdot \ ] $ that do mix. One-loop transitions from  $\ab{ \ \cdot \ }^3$-type operators to  $\ab{ \ \cdot \ }[ \ \cdot \ ]$ is forbidden by helicity selection rules. These patterns are summarised by the black arrows in the figure.   With blue arrows we indicate the two-loop  anomalous dimensions that are computed in the sections indicated on top. 
	}
	\label{figdiag}
\end{figure}

While chunks of the   SM  dimensions-six operators' anomalous dimensions have been computed in many papers in the literature, see  e.g. \cite{Grojean:2013kd,Hagiwara:1993ck,Hagiwara:1994pw,Alam:1997nk,Mebane:2013zga,Chen:2013kfa,Elias-Miro:2013gya} and  \cite{Elias-Miro:2013mua,Elias-Miro:2013eta} for the  anomalous dimensions    relevant  for Higgs and Electroweak physics,  the full matrix has been computed in  \cite{Jenkins:2013zja,Alonso:2013hga,Jenkins:2013wua,Alonso:2014zka}. 
With this we conclude the summary of the one-loop pattern, see figure \ref{figdiag} for an outline.

A particularly interesting class of two-loop anomalous dimensions consists of  those RG mixings that are forbidden at one-loop.    We distinguish two types of such two-loop  RG mixings.
The first type are those two loop RG mixings that decrease  the number of particles by a unit, namely  an operator with $\l$ legs can mix, at two loops,  with an operator with $\l-1$ legs.  These transitions  occur in the direction opposite to the long 1-loop arrow in figure \ref{figdiag}.
The second type are those two loop RG mixings that are forbidden at one loop solely by selection rules (but not by leg counting). Those would be two loop transitions from $\ab{ \ \cdot \ }^3 \rightarrow  \ab{ \ \cdot \ } [ \ \cdot \ ]$ and  two loop transitions among the four-particle operators of the classes  $ \ab{ \ \cdot \ } [ \ \cdot \ ]$ and $ \ab{ \ \cdot \ }^2$.

In this work we will investigate the two-loop anomalous dimensions indicated by blue arrows in figure \ref{figdiag}.  These  anomalous dimensions arise from  the  following type  of diagrams
\be
\begin{minipage}[h][1.5cm][t]{0.19\linewidth} \vspace{.25cm} \twoloopgenUn \end{minipage}  \ +  \  \ 
\begin{minipage}[h][1.5cm][t]{0.19\linewidth} \vspace{.25cm}  \twoloopgenDos \end{minipage}    \, ,  \label{gentwoloop}
\ee
where the  on-shell particle lines denote   any allowed  SM particle and the scattering element of each term is evaluated at tree-level. 
Indeed, under the   three-particle cut of the first term of \reef{gentwoloop},   three particles  are removed from the operator $O_i$ while the $S_{2\leftarrow 3}$ scattering matrix element produces two particles. By means of this process  
  an operator with $\l$ legs  mixes, at two loops,  with an operator with $\l-1$ legs.
There are no 
 contributions that involve  a 1-loop (or higher loop) scattering matrix element.  Indeed, at two-loop order in the anomalous dimension matrix, such 1-loop scattering matrix element can only be contracted with two particles of the FF and thus does not decrease the number of particles in $O_i$. Finally, in principle we can    also reduce the number of legs in the operator by closing the particles of the FF into a loop, this is indicated by the second term in \reef{gentwoloop}.
However,  in the next section we will show that the second term  in \reef{gentwoloop}, involving the one-loop FF, vanishes  in the two-loop mixing transitions that decrease the number of legs -- the disappearance of this term was noted before in ref.~\cite{Bern:2019wie}.

In conclusion, for this class of anomalous dimensions, the only  contribution  comes from a tree-level S-matrix  element, a tree-level FF and a  three-particle cut (first term  in \reef{gentwoloop}). 
They can be efficiently calculated and therefore  the on-shell method is  well suited  for computing the leading RG mixings that shorten the length of the operator.

\section{The `easy' two-loop anomalous dimensions}
\label{easyQ}

In the following we  compute the two-loop anomalous dimensions of the RG mixing shown by blue arrows in figure \ref{figdiag}. 
This is a higher loop order calculation of the r.h.s. of \reef{chw1} than what we have encountered so far. However it is still order ``$(0)$"  because,  as we will see,  such two-loop calculations lead to a Lorentz invariant polynomial in the spinors $\lambda$, $\tilde \lambda$. Thus, according to the l.h.s. of \reef{chw1},   its coefficient should be interpreted  as the anomalous  dimension $\gamma$ because  $\mu\partial_\mu$ picks  up the coefficient  multiplying a single leading logarithm.~\footnote{Had the one-loop not vanished, we would have needed to expand $e^{-i\pi D}$ to  second order to get rid of the leading $\log^2$ term and disentangled the coefficient of the one-loop  and the two-loop single log.}

The contributions coming from the first diagram in \reef{gentwoloop}, consisting in $(-1/\pi)$ times the three-particle phase space integral of a matrix element and a form factor, can be written as
 \be
\frac{\langle 12\rangle[12]}{(16\pi^2)^2 3!}\int d\Omega_3 \, M(12 ; 1^\prime 2^\prime 3^\prime) \, F_{O_i}(1^\prime 2^\prime 3^\prime 4 \dots n)  \, ,  \label{rot00}
\ee
where   the spinors in the integrand are written in terms the two external spinor by the following rotation
\be
 \begin{pmatrix}
	\lambda^\prime_1 \\
	\lambda^\prime_2 \\
	\lambda^\prime_3 
\end{pmatrix} = 
\begin{pmatrix}
	\cos {\th_1} & -e^{i\phi}\cos {\th_3}\sin {\th_1} \\
	\cos {\th_2}\sin {\th_1} & e^{i\phi}\left(\cos {\th_1}\cos {\th_2}\cos {\th_3}-e^{i\delta}\sin {\th_2}\sin {\th_3}\right) \\
	\sin {\th_1}\sin {\th_2} & e^{i\phi}\left(\cos {\th_1}\cos {\th_3}\sin {\th_2}+e^{i\delta}\cos {\th_2}\sin {\th_3}\right)
\end{pmatrix}
\begin{pmatrix}
	\lambda_1\\
	\lambda_2
\end{pmatrix} \label{rot3} \, , 
\ee
and the  measure   is $
 d\Omega_3 = 4\cos\th_1\sin^3\theta_1d\th_1 2\cos\th_2\sin\th_2d\th_22\cos\th_3\sin\th_3 d\th_3\,\frac{d\delta}{2\pi}\,\frac{d\phi}{2\pi}$ \cite{Zwiebel:2011bx,Caron-Huot:2016cwu}.
 The $1/3!$ in \reef{rot00} arises only when considering the phase space for three identical particles.  Further details on the phase-space integral can be found in appendix \ref{psi}.

\subsection{$ \ab{ \ \cdot \ }^2 \, \rightarrow \,  \ab{ \ \cdot \ }^3 $:  4-point to 3-point}
\label{case1}

We start by computing the transitions from four-field operators into three-field operators. 
There are only two structurally different contributions at two loops, namely the $\phi^2 F^2\to F^3$ and the $\psi^2 \phi F\to F^3$ mixings. We start with the former contribution.
In the SM this transition is only possible when the gauge field belongs to the $SU(2)_L$, thus we compute the RG mixing from  the operator   
$O_{WW}$
to  $O_{3W}$, see table \ref{tabbb} with $F=W$. The corresponding MFFs are 
\bea\nn
&F_{3W}(1^-_A 2^-_B 3^-_C)= \frac{i}{\sqrt{2}}f^{ABC}\ab{12}\ab{23}\ab{31}  \\
&F_{WW}(1_i 2^-_A 3^-_B 4^*_j)= -2 \, \delta_{i}^{\, j}\delta^{AB}\, \ab{23}^2.
\label{eq:mff3WandWW}
\eea
We find the following potential contributions \vspace{-.1cm}
\be
\begin{minipage}[h][1.75cm][t]{0.21\linewidth} \vspace{.2cm} \dgenUnP{boson}{boson}{scalar}{scalar}{boson}{scalar}{scalar}{boson}{boson} \end{minipage}     +  
\begin{minipage}[h][1.75cm][t]{0.22\linewidth} \vspace{.2cm}  \dgenDosP{boson}{boson}{boson}{boson}{boson}{scalar}{scalar}{boson}{boson} \end{minipage}  + 
\begin{minipage}[h][1.75cm][t]{0.22\linewidth} \vspace{.2cm}  \dgenDosPP{boson}{boson}{boson}{boson}{boson}{scalar}{scalar}{boson}{boson} \end{minipage}
\, .  \label{genm1} \vspace{-.1cm}
\ee
We have excluded  one loop FFs that are contracted with the amplitude $H_i H_j \rightarrow W^- W^-$, because this process is zero at tree-level in the SM.
Next, note that the one-loop FFs in \reef{genm1} vanish. Indeed,  the loop is given by 
$
\int d^d \l \frac{\l^\mu}{[(\l+p/2)^2-i \eps][(\l-p/2)^2  -i\eps ]}=0$.~\footnote{The same integral appeared recently in   \cite{Bern:2019wie} when computing the two-loop mixing of $O_6$ into $O_{3W}$.} 
 %After performing the change of variables $l=-k$ one has $I(p)=-I(-p)$ and therefore the integral vanishes. Then,  introduce Feynman parameters and a small IR regulator $\eps>0$ to find $  \int d^d \l  \int_0^1 dx\frac{\l^\mu}{(\l^2 +\l\cdot p (1-2x) + \eps)^2} $, where we have used that $p^2=0$ on-shell.  Changing variables $\l=k-p(1-2x)/2$ and integrating over $x$, we get    $\int d^d k \frac{k^\mu}{(k^2+\eps)^2}=0$, since the integrand is odd.
  We are left with the first term in \reef{genm1}, involving a three-particle phase-space integration between a tree-level matrix element and a minimal FF. 
The five-point amplitude  is given by 
\be
M(1_i 2^-_{A_2} 3^-_{A_3} 4^+_{A_4} 5^*_j) \,=\, 2\sqrt{2} g^3  [14]^2[45]^2\, \left( \frac{ (T^{A_2}T^{A_3}T^{A_4})_{ij}}{[12][23][34][45][51]} + \text{perms. \{2,3,4\}}   \right)  \,,  \label{fiveamp}
\ee
which has to be contracted with
the  MFF in  \reef{eq:mff3WandWW},   \vspace{-.1cm}
\be
\begin{minipage}[h][1.75cm][t]{0.21\linewidth} \vspace{.2cm} \dgenUnP{boson}{boson}{scalar}{scalar}{boson}{scalar}{scalar}{boson}{boson} \end{minipage} =  3 \frac{\ab{12}[12]}{(16\pi^2)^2} 
\int d\Omega_3 M(y_i 1^-_A 2^-_B x^+_\mathsf{x} z^*_j)  \, F_{WW}(y_i 3^-_C x^-_\mathsf{x} z^*_j)  \, ,  \label{tttccc}   \vspace{-.1cm}
\ee
where we denoted by $x,\,y$ and $z$ the momenta of the spinors we are integrating over. 
In  \reef{tttccc} we   included a factor of $3$ that accounts for all possible ways to contract the external state $\langle 1^-_A 2_B^- 3_C^- |$ with the $M\otimes F$.
 The integral is done by rotating
$\lambda_x, \lambda_y,\lambda_z$ in terms of $\lambda_1,\lambda_2$ as in \reef{rot3}. 
The integrand of \reef{tttccc}  involves  permutations of 
$[ijk] \equiv   \frac{[xy]^2[xz]^2\ab{x3}^3}{[yi][ij][jk][kz][zy]}$, with $i,j,k=1,2,x$, multiplied by the corresponding trace of $T^{A}$'s matrices, generated from the contraction of the color structure in \reef{fiveamp} with the $\delta_{ij}$ of the $F_{WW}$ form factor. 
The integral measure is invariant  under $y\leftrightarrow z$, which is implemented with the change of variables $(\th_3, \delta)\rightarrow (\pi/2-\theta_3,\delta+\pi)$. Therefore, we can reverse the ordering of the arguments to get $[ijk]=- [kji]$, and write the integrands proportional to $\tr(T^AT^BT^C)$ in terms of the ones proportional to  $\tr(T^CT^BT^A)$.
Thus we are lead to 
\be
  \reef{tttccc}= -\frac{\ab{12}[12]}{(16\pi^2)^2} \, 6\sqrt{2}g^3  i f^{ABC}
\int d\Omega_3\, \big(  [12x] +  [x12] +[2x1] \big)
\label{hhfftofff_eq} \, , 
\ee
where we have used  $\tr(T^A[T^B,T^C])=i f^{ABC}/2$ and a factor of $3$ comes from the permutations in \reef{tttccc}.  
The angular integrals in \reef{hhfftofff_eq} are given by
\be
\int d\Omega_3  \, [12x] =  \int d\Omega_3 \,  [x12] = \frac{1}{2}\frac{\ab{13}\ab{23}}{[12]} \ ,\quad
\int d\Omega_3 \,  [2x1] = \frac{3}{2}\frac{\ab{13}\ab{23}}{[12]} \, . 
\ee
From this we read  the two loop anomalous dimension
\be
\gamma_{3W\leftarrow WW}\,=\,  30\frac{g^3}{(16\pi^2)^2}  \,  . 
\ee

Next we compute  a transition of the type $\psi^2 \phi F \rightarrow F^3$. 
In particular we compute  the renormalisation of the pure glue operator $O_{3G}$ by the gluon dipole $O_{qG}$.  This is the opposite mixing direction that we  have computed in  section \ref{nonMFF}. The MFFs are already given in \reef{FF3G}.
There are the following possible contributions
\be
\begin{minipage}[h][1.75cm][t]{0.21\linewidth} \vspace{.2cm}  \dgenUnP{boson}{black}{scalar}{black}{black}{scalar}{black}{black}{boson} \end{minipage}    +  
\begin{minipage}[h][1.75cm][t]{0.22\linewidth} \vspace{.2cm}  \dgenDosP{boson}{boson}{scalar}{boson}{scalar}{black}{black}{boson}{boson} \end{minipage}  + 
\begin{minipage}[h][1.75cm][t]{0.22\linewidth} \vspace{.2cm}  \dgenDosPP{boson}{boson}{scalar}{boson}{scalar}{black}{black}{boson}{boson} \end{minipage}
 \, ,  \label{gen0}
\ee
However, the one loop FF vanishes because, due to the presence of $\sigma_{\mu\nu}$ in the dipole vertex, the loop of fermions involves the trace of an odd number of $\gamma^\mu$-matrices. 
We have excluded  one-loop FFs that are contracted with the  amplitudes  $\psi^+\psi^+\rightarrow g^- g^-$ or $\phi\phi \rightarrow g^-g^-$ which do not exist, at tree-level,  in the SM Lagrangian.
Again, the contribution reduces to compute the  first term in \reef{gen0} which involves a tree-level matrix element and a minimal FF. 
The five-point scattering amplitude has already appeared  in \reef{ampffsgg}. Since in the current case the $SU(2)$ indices are relevant, they are denoted by denoted by $i,j$, while the color ones by $\alpha,\beta$. Both the amplitude and FF proportional to the trivial factor $\delta_{ij}$.
The convolution of the amplitude and the MFF is
\be 
\int \text{d}\Omega_3  \, M(y^+_{\alpha i} 1^-_A 2^-_B z^+_\beta x_j)\, F_{qG}(y^-_{\alpha i} 3^-_C z^-_\beta x^*_j) = -4\sqrt{2}yg_s^2 if^{ABC} \int\text{d}\Omega_3  \frac{ [yz]^2 \ab{y3} \ab{3z}}{[y1][12][2z]} 
\ee
where we used the fact that the integral measure is invariant under $y\leftrightarrow z$ to write the two color orderings in a similar way, and that $ i f^{ABC}/2= \tr(T^A[T^B,T^C])$ and $\delta_{ij}\delta_{ij}=2$. The integral over the phase space angles is straightforward to do and gives $\frac{1}{2}\ab{13}\ab{23}/[12]$. Finally, taking into account a factor of $3$ for the  permutations of the three different  gauge bosons, we get
\be
\gamma_{3G\leftarrow qG}  =  12\frac{yg_s^2}{(16\pi^2)^2}.
\ee
This computation, compared with the one loop reverse transition in  \ref{nonMFF}, shows explicitly that the complexity of the computations in this method does not scale strongly with the number of loops, and two-loop computations can be easier than one-loop ones.

These two Lorentz structures, $\phi^2F^2$ and $\psi \phi F$, are the only four-particle operators that mix with the three-particle structure $F^3$ at two loops with a single log.

\subsection{$ \ab{ \ \cdot \ }  \, \rightarrow  \,  \ab{ \ \cdot \ }^2$:  5-point to 4-point}
\label{case2}

We start by computing a transition of the type  $\psi^2\phi^3 \rightarrow  \psi^2 \phi F$.
For concreteness we compute the renormalisation of the hypercharge dipole of the leptons $\mathcal{O}_{eB} = \bar{\ell}_L \sigma_{\mu\nu} e_R HB_{\mu\nu}+h.c.$ due to the leptonic Yukawa $\mathcal{O}_{y_e}=|H|^2\bar{\ell}_L e_RH +h.c.$, with other pairings being similar since the non-abelian part of the amplitude plays no role in the computation. 
The potential contributions to the $\psi^2\phi^3 \rightarrow  \psi^2 \phi F$ mixing are given by
\be
\begin{minipage}[h][1.5cm][t]{0.16\linewidth} \vspace{.25cm} \dgenUn{boson}{black}{scalar}{scalar}{black}{scalar}{scalar}{black}{scalar}{black} \end{minipage}   +   
\begin{minipage}[h][1.5cm][t]{0.16\linewidth} \vspace{.25cm} \dgenUn{boson}{scalar}{scalar}{scalar}{scalar}{scalar}{scalar}{scalar}{black}{black} \end{minipage}  +  
\begin{minipage}[h][1.5cm][t]{0.18\linewidth} \vspace{.25cm}  \dgenDos{black}{boson}{black}{boson}{black}{scalar}{scalar}{boson}{black}{scalar} \end{minipage}  \, .  \label{gen1}
\ee
The second contribution with the three scalar in the cut is zero. This can be seen by explicit computation, but also by noticing that the fermion structure $\ab{12}$ of the Yukawa is unaltered by the integration and there is no dimension-six invariant with the $F\psi^2 \phi$ particle content  and  proportional to the fermion spinors $\ab{12}$.
The third contribution also vanishes, because the FF loop vanishes as we showed in the previous section.
We are left with the tree-level amplitude contribution
\be
\begin{minipage}[h][1.5cm][t]{0.21\linewidth} \vspace{.05cm} \dOytoOdDos \end{minipage} = \frac{\ab{14}[14]}{(16\pi^2)^2}\int d\Omega_3  \,  M(1_i^-   x^+_a y_b^* z_c  4^-)  \,  F_y(x^-_a 2^- 3_j  y_b  z^*_c) \, , \label{d2l}
\ee
The amplitude  in \reef{d2l} is given by
\be
M(1_i^- 2_j^+ 3_k 4^*_l  5^-)  =  - 2 \sqrt{2}  g^{\prime}  (g^{\prime 2}Y_L Y_H\delta_{ij}\delta_{kl}+g^2T^A_{ij}T^A_{kl})\left( Y_H \frac{[23][24] }{[12][35][45]} - Y_L\frac{[23][24] }{  [15][25] [34]}\right) \, ,  \label{amtoint}
\ee
where we included both $U(1)_Y$ and $SU(2)_L$ contributions to the amplitude, $Y_L$ refers to the  lepton doublet and $Y_H$  the Higgs doublet Hypercharge. 
In first term in the parenthesis of \reef{amtoint}   the $U(1)_Y$ boson $5^-$ is attached to the scalars, while in the second term the  gauge boson is attached to the fermions. 
The  FF in \reef{d2l} is 
\be
F_{y_e}(1^-_i 2^- 3_j 4^*_k 5_l) = \ab{12} \, (\delta_{ij}\delta_{kl}+\delta_{il}\delta_{kj})
\ee
Next, the spinors associated to the particles $x,y,z$ are rotated as in  \reef{rot3} leaving a simple trigonometric integral that gives
\be
\int\text{d}\Omega_3 \left( Y_H \frac{[xy][xz]\ab{x2} }{[1x][y5][z5]} - Y_L\frac{[xy][xz]\ab{x2} }{  [15][x5] [yz]}\right) \,=\, -Y_H \frac{\ab{42}}{[14]} 
\ee
with the term proportional to the fermion hypercharge vanishing since the integrand is odd under $y\leftrightarrow z$ while the integral measure is invariant.
Therefore all in all we get
\be
\reef{d2l} = \frac{1}{(16\pi^2)^2} g^\prime Y_H \left(g^{\prime 2}Y_LY_H(N+1)+g^2\frac{N^2-1}{2N}\right)2\sqrt{2}\delta_{ij} \,\ab{14}\ab{42} \, , \label{antt}
\ee
where $N=2$ for the doublet. 
Exchanging $1^-\leftrightarrow 2^-$ in  \reef{d2l} amounts to $Y_L\leftrightarrow Y_R$ in the order $g^{\prime 3}$ contribution. Thus all in all,  we  get the anomalous dimension  
\be
\gamma_{eB\leftarrow y_e} = \frac{3}{(16\pi^2)^2} \left(g^{\prime 3}(Y_L+Y_R)Y_H^2 +g^\prime g^2 Y_H\frac{ 1}{4}\right),
\ee
in  agreement  with \cite{Panico:2018hal}.

The second case we consider is the renormalisation of the $\phi^2F^2$ operators due to the dimension-six Yukawa $O_y$. For instance, take the operator to be $O_{GG} $.  As before, denote by $\alpha,\beta$ the color indices and by $i,j,...$ the $SU(2)_L$ ones.
There are two types of contributions, given by the diagrams
\be
\begin{minipage}[h][1.5cm][t]{0.16\linewidth} \vspace{.25cm} \dgenUn{boson}{boson}{scalar}{black}{black}{scalar}{black}{black}{scalar}{scalar} \end{minipage}   +  \, 
\begin{minipage}[h][1.5cm][t]{0.17\linewidth} \vspace{.25cm}  \dgenDos{boson}{boson}{scalar}{boson}{scalar}{black}{black}{boson}{scalar}{scalar} \end{minipage}  
\, .  \label{gen2}
\ee
The one-loop FF vanishes as before, and we have excluded those one-loop FFs that are contracted with $M(\phi\phi\rightarrow g^-g^-)=0$ at tree-level. 
Thus we are left with the first contribution in \reef{gen2}. Both the  MFF and the 5-point amplitude have already appeared and  they are given in   \reef{OyMFF} and  \reef{ampffsgg}, respectively. The phase-space integral in  \reef{ddcite}  gives 
\be
\int\text{d}\Omega_3  \, M(y^+_{\alpha k} 1^-_A 2^-_B z^+_{\beta} x^*_l)  \, F(y^-_{\alpha k} z^-_{\beta j} x_l 3_i 4^*_j) =  - 
y g_s^2 \delta^{AB}\delta_{ij}(N+1)\,2\int \text{d}\Omega_3 \frac{[yz]^2\ab{yz}}{[y1][12][2z]} \, , 
\ee
where the factor 2 comes from the two color orderings giving the same integrand, $\delta^{AB}=2\tr(T^AT^B)$ and the $(N+1)$ factor, with $N=2$, comes from the $SU(2)_L$ contraction. The integral over the trigonometric angles gives $-\ab{12}/[12]$. The result is  
\be
\gamma_{FF\leftarrow y} = -3\frac{yg_s^2}{(16\pi^2)^2} \, . 
\ee

The last contribution of the type $\ab{ \ \cdot \ } \rightarrow \ab{ \ \cdot \ }^2$ is the $\psi^2\phi^3 \rightarrow  \psi^4$ transition, 
\be
\begin{minipage}[h][1.5cm][t]{0.22\linewidth} \vspace{.05cm} \dOytoOpsipsi \end{minipage}  
 \label{ddcite} \, .
\ee
There is not a contribution from a one-loop FF. Note that the  one-loop form factor where one scalar line  is closed into a fermion leg  vanishes on-shell because it is proportional to $\cancel{\partial} q$. Indeed such loops are proportional to $(\bar q \cancel{\partial} q)|H|^2+h.c.$ in an effective action calculation, see e.g. \cite{Elias-Miro:2013mua}.

For definitenes, we compute the renormalisation of the 4-fermion operator $O_{lequ} = \epsilon^{ij}(\bar{\ell}_L^i e_R)(\bar{q}^j u_R)$ by the up-quark Yukawa $O_{y_u} = |H|^2\bar{q}_L^i u_R \epsilon_{ij}H^{j*}$ with MFF
\bea
F_{lequ}(1^-_i2^-3_j^-4^-) &= \epsilon^{ij} \ab{12}\ab{34}    \, , \\[.2cm]
F_{y_u}(1^-_i 2^- 3^*_j 4_k 5^*_l) &= (\epsilon_{ij}\delta_{kl}+\epsilon_{il}\delta_{kj}) \,\ab{12} \, . 
\eea
Any other choice of Yukawa and 4-fermion operators will be computed in a similar fashion. The 5-point amplitude involved has three components, proportional to $y_e^3$, $y_e\lambda$ and $y_e g^2$ or $y_e g^{\prime2}$. 
The contributions dependent on the gauge coupling are proportional to the difference of two scalar momenta, say $(p_5-p_4)_\mu$, since there is always a scalar current coupled to a gauge boson. Since the FF is independent of the scalar momenta, the phase space integration sets these contributions to zero. The only relevant part of the amplitude is then
\be
M(1^-_i 2^- 3_j 4^*_k 5_l) = - y_e^3 \,\left( \delta_{i}^{\, j}\delta_{k}^{\, l}\frac{[35]}{ [13] [25]}+
\delta_{i}^{\, l}\delta_{k}^{\, j}\frac{[53]}{ [15] [23]}\right)\,+\, 2y\lambda (\delta_{i}^{\, j}\delta_{k}^{\, l}+\delta_{i}^{\, l}\delta_{k}^{\, j}) \frac{1}{[12]}  \, . 
\ee
The convolution is 
\be 
\int\text{d}\Omega_3 M(\ell^-_i e^- x_a y^*_b z_c)F_{y_u}(q^-_j u^- x^*_a y_b z^*_c)  =
2 (N+1) \epsilon_{ji}\, \int\text{d}\Omega_3\left[ - y^3_e \frac{[xz]}{[\ell x][e z]}\ab{q u} +2 y_e\lambda \frac{\ab{q u}}{[\ell e]}   \right]  \, , 
\ee
where we labeled the external momenta as $\ell, e, q, u$ to keep track of the fermions' flavour, and $N=2$ is the dimension of the doublet. The second trigonometric integral is trivial and gives 1, while the first one gives $2\ab{q u}/[\ell e]$. After multiplying this by the phase space factor $\ab{\ell e}[\ell e]$ and a $1/2!$ for the identical particles $x$ and $z$, we identify the resulting factor $\epsilon^{ij}\ab{\ell e}\ab{qu}$ as the one associated with the operator $O_{lequ}=\epsilon^{ij}(\bar{\ell}^i_L e_R)(\bar{q}^j_L u_R)$, with an anomalous dimension 
\be
\gamma_{lequ\leftarrow y_u} \,=\,   -  \frac{6}{(16\pi^2)^2} \,(  y_e \lambda -  y_e^3  ) \, .
\ee

\subsection{$1 \,  \rightarrow  \,  \ab{ \ \cdot \ } $:  6-point to 5-point}
\label{case3}

Lastly, consider the transition between  the operator $O_6=|H|^6$ with MFF
\be
F_6(1_i 2^*_j 3_k 4^*_l 5_m 6^*_n) =  6\, \delta_{i}^{\,j}\delta_{k}^{\, l}\delta_{m}^{\, n} + \text{5 perm. of $\{j,l,n\}$}  
\ee
and the dimension-six Yukawa $O_y$. The only non-vanishing contribution comes from the diagram
\be
\begin{minipage}[h][1.5cm][t]{0.2\linewidth} \vspace{.05cm} \dOsixtoOy  \end{minipage}   \label{ddcite} \, .
\ee
The integral over the phase space  is the same we encountered  in the  preceding section for the $\psi^2\phi^3\to \psi^4$ transition. The flavour contraction gives $2(N+2)(\delta_{i}^{\, j}\delta_{k}^{\, l}+\delta_{i}^{\, l}\delta_{k}^{\, j})$. Thus we are led to
\be
\gamma_{y_e \leftarrow 6} \,=\,   \frac{48}{(16\pi^2)^2} (\, y_e\lambda  -y_e^3 \,) \, .
\ee

\section{Conclusions}
\label{conc}

In this work we have demonstrated how to retrieve the anomalous dimensions of the SM EFT operators using the  on-shell  S-matrix and form factors.
In particular we showed how only tree-level S-matrix elements and FFs can be used to determine all of the one-loop and many two-loop anomalous dimensions of the SM EFT. 
 Another  nice aspect of this method is that it re-uses the same amplitudes  to determine many anomalous dimensions. For instance,   the amplitude 
$M(FF \phi \psi \psi)$ appears in the calculation of the one loop $F^3 \rightarrow \psi^2 \phi F $, $  \phi^2 F^2 \rightarrow \phi^3 \psi^2   $,  and the two loop $ \psi^2 \phi^3 \rightarrow \phi^2 F^2 $  and  $   \psi^2 \phi F  \rightarrow F^3 $ mixings. This aspect of the method permits many self-consistency checks. 
Our calculations  could  be fully  automated and extended to the entire one-loop anomalous dimension matrix of the leading higher dimensional operators in the SM EFT, or a more general field theory.

In section \ref{oneloopxcheck} we computed a variety of representative one-loop anomalous dimensions including  
 the order $\lambda$ corrections,   examples involving  IR  anomalous dimensions  and  non-minimal form factors. 
In section  \ref{recap} we have  discussed  the structure of the one-loop anomalous dimension matrix of dimension-six operators. This has allowed us to identify   the most interesting two-loop anomalous dimensions, namely those   that are forbidden at one-loop either by leg counting  
 or by  helicity selection rules only. 
 In section \ref{easyQ} we have computed the two-loop anomalous dimensions 
that mix operators as indicated by the blue arrows in figure \ref{figdiag}
as an illustration of  the power of the on-shell method.

Current and future precision experiments probe and will probe the RG structure at two loops. The paradigmatic example is the measurement of the  electron EDM \cite{Andreev:2018ayy,Panico:2018hal}, whose current bound tests multi-TeV dynamics at two loops. 
Other future experiments, like the  measurements of the $\text{Br}(\mu\to e\gamma)$ at the MEG-II experiment will reach a precision of $6\cdot 10^{-14}$ \cite{Baldini:2018nnn}. This would translate into a bound of $\Lambda\gtrsim 10^3\,\text{TeV}$ for the operator $\mathcal{L}\supset \frac{\sqrt{y_ey_\mu}}{\Lambda^2}\bar{\mu}\sigma_{\mu\nu}eHF_{\mu\nu}$. If the UV dynamics is such that, instead of the dipole, an operator $\frac{\sqrt{y_ey_\mu}}{\Lambda_i^2} O_i$, which only affects the dipole at two loop order, is generated at the high scale, then the log-enhanced contribution due to the two-loop RG mixing would set a constraint of $\Lambda\gtrsim 10\,\text{TeV}$.
Having a complete map of the two-loop anomalous dimensions of the operators probed by this and other future precision experiments, will provide the complete characterisation of the dynamics affecting such processes through a log-enhanced contribution.

While we have shown how to compute  the transitions
$ \ab{ \ \cdot \ }^2 \, \rightarrow \,  \ab{ \ \cdot \ }^3 $, 
$ \ab{ \ \cdot \ }  \, \rightarrow  \,  \ab{ \ \cdot \ }^2$ and  
 $1 \,  \rightarrow  \,  \ab{ \ \cdot \ } $, it will be interesting to compute the rest of the two-loop anomalous dimensions for which the corresponding one-loop contributions are absent,  as well as  those for which the one-loop contribution are suppressed by a small coupling, like a Yukawa of a light particle. 
In addition,  we  look forward to investigating other aspects of this method. Whether we can  relate   the  branch-cut discontinuity of other physical quantities to the Callan-Symanzik  equation using the dilatation operator \reef{genD},  or compute certain anomalous dimensions at large orders, are some of the questions that  we are eager to investigate further.

 \noindent
\textbf{Note added:} Simultaneously with the submission of this preprint Ref.~\cite{Baratella:2020lzz} appeared where   the renormalisation of higher-dimensional operators from on-shell amplitudes is presented from a slightly different perspective.  
In Ref.~\cite{Baratella:2020lzz}  all one-loop anomalous dimensions of the $SU(2)_L$ dipole operator of the electron (up to self-renormalisation) are reproduced.

\noindent  Shortly after, Ref.~\cite{1797240} appeared where  the computation of SM EFT operator anomalous dimensions using unitarity cuts is also discussed and   the self-renormalisation of  $|H|^2 B_{\mu\nu}^2$   and $G_{\mu\nu}^3$  is reproduced.  There instances of applying these techniques to  operators of dimension  larger than six are also discussed. 

 \noindent Before we submitted to the journal,  Ref.~\cite{Bern:2020ikv} appeared where they also describe the use of on-shell methods  to compute one and two-loop anomalous dimensions in the context of EFTs containing higher-dimension operators.
There the  pattern of vanishing entries in the two-loop anomalous dimension matrix is also analysed.

\subsection*{Acknowledgements}

JEM and JI  are grateful to the participants of the  SISSA-ICTP joint Journal Club for useful discussions on this topic. 
MR acknowledges funding from the Swiss National Science Foundation under grant no. PP002-170578.

%%%%%%%%%%%%%%%%%%%%%%%%%
%%%%%%%%                           %%%%%%%%%
%%%%%%%%   Appendix    %%%%%%%%%%%
%%%%%%%%                           %%%%%%%%%
%%%%%%%%%%%%%%%%%%%%%%%%%

\appendix

\section{Phase Space Integrals}
\label{psi}

We are interested in evaluating ``$n\rightarrow2$'' integrals of the type  
\bea
I_{n\rightarrow2}=\int\frac{1}{n!}\prod_{i=1}^{n}\frac{d^3p'_i}{(2\pi)^3\, 2p_i'^0}\,\,(2\pi)^4\delta^4\left(p_1+p_2-\sum_{i=1}^{n}p'_i\right)\matrixel{12}{  M_{n\rightarrow2}}{1'\dots n'}\matrixel{1'\dots n'}{{\cal O}}{0}
\eea 
These integrals feature an S-matrix element with $n$ particles in the initial state and 2 in the final state. We include a symmetry factor $1/n!$ for identical particles.  

For massless particles, the integrand is expressed more conveniently in terms of spinor helicity variables $\lambda'_i$, rather than the momenta.  
Assuming that $s_{12}=\langle12\rangle[21]$ is nonzero, the two spinors corresponding to the external particles $\lambda_1$ and $\lambda_2$ are linearly independent, and form a complete basis. Therefore the spinors for the $n$ exchanged particles (whose momenta are integrated over) may be expressed as
\begin{equation}
	\begin{pmatrix}
		\lambda'_1 \\ \lambda'_2 \\ \vdots \\ \lambda'_n
	\end{pmatrix} = A
	\begin{pmatrix}
		\lambda_1\\ \lambda_2
	\end{pmatrix}, \qquad \qquad A = 
	\begin{pmatrix}
		a_{11} & a_{12} \\ a_{21} & a_{22} \\ \vdots & \vdots \\ a_{n1} & a_{n2}
	\end{pmatrix},\label{Eq:basisch}
\end{equation}
where the $a_{ib}$ are arbitrary complex numbers.
Next we  recast the measure and momentum conserving delta function  according to the change of variables in  \reef{Eq:basisch}.

The momentum conserving delta function  is then  given by 
\be
\delta^4(P)=4\delta(\lambda^1\bar{\lambda}^{\dot{1}})\delta(\lambda^1\bar{\lambda}^{\dot{2}})\delta(\lambda^2\bar{\lambda}^{\dot{1}})\delta(\lambda^2\bar{\lambda}^{\dot{2}})  =  \frac{4}{\langle 12\rangle^2[12]^2}\delta(u_{11})\delta(u_{12})\delta(u_{21})\delta(u_{22}) \, ,  \label{dd1}
\ee
where
$
P^{\alpha\dot\a}\equiv \lambda^\alpha\bar{\lambda}^{\dot{\alpha}}\equiv\lambda^\alpha_1\bar{\lambda}^{\dot{\alpha}}_1 + \lambda^\alpha_2\bar{\lambda}^{\dot{\alpha}}_2 - \sum_{i=1}^{n}\lambda_i^{\prime\alpha}\bar{\lambda}_i^{\prime\dot{\alpha}}
$.
And we have defined the $2\times 2$ matrix $u_{nm}= \delta_{nm}- (A^\dagger A)_{nm}$. 
Thus, the  momentum conserving delta function imposes the constraint 
 $
A^\dagger A = \mathds{1}_2
 $.

Now we consider the measure for one of the $n$ exchanged particles. We would like to express it in terms of the $a_{ib}$ variables, perhaps like:
\be
\frac{d^3p'_i}{(2\pi)^3\,2p_i^{\prime 0}} \stackrel{?}{=} f(a_{i1},a_{i2})da_{i1}da^*_{i1}da_{i2}da^*_{i2}.
\ee
This measure only depends on $a_{i1}$ and $a_{i2}$ because the measure depends only upon $p_i^{\prime\mu}$ and $p_i^{\prime\mu}\sigma^{\alpha\dot{\alpha}}_\mu = \lambda_i^{\prime\alpha}\bar{\lambda}_i^{\prime\dot{\alpha}}$. However, there are three real variables to be integrated over on the LHS but four on the RHS, so the relationship above cannot be true.
The problem arises due to a redundant phase that emerges when the kinematic information is expressed in the $a_{ib}$ variables rather than in the momenta. To be specific, we can rotate complex $a_{i1}$ and $a_{i2}$ by a common phase so that
\be
a_{i1}\rightarrow e^{i\a_i}a_{i1}, \quad a_{i2}\rightarrow e^{i\a_i}a_{i2}, \quad \lambda'_i \rightarrow e^{i\a_i} \lambda'_i.
\ee
This rotation leaves $p_i^{\prime\alpha\dot{\alpha}} = \lambda_i^{\prime\alpha}\bar{\lambda}_i^{\prime\dot{\alpha}}$ unchanged, so clearly the measure, and the momentum conserving delta function are invariant  under this rotation.
Equally, the integrand is also  independent of this global phase
\be
\matrixel{12}{{\cal M}_{n\rightarrow2}}{1'\dots n'}\matrixel{1'\dots n'}{{\cal O}}{0} \rightarrow e^{2ih_i\a_i}\matrixel{12}{{\cal M}_{n\rightarrow2}}{1'\dots n'}\cdot e^{-2ih_i\a_i}\matrixel{1'\dots n'}{{\cal O}}{0},
\ee
where $h_i$ is the helicity of particle $i$.
All quantities in the integral are therefore invariant under the global phase transformation, so that adjusting it is analogous to performing a gauge transformation in a gauge theory. Just as in gauge theory, it is convenient to fix a value for this ``gauge parameter''. This can be done in a way that still leaves the equivalence of different gauge fixing conditions manifest using the Fadeev--Popov trick: we multiply the measure by the following factor 
\be
\int d\a_i\delta\left(f_i(\a_i)\right)f_i^\prime(\a_i) = 1.
\ee
Defining the overall phase to be $\a_i\equiv\arg(a_{i1})+\arg(a_{i2})$, the measure becomes
\be
	\frac{d^3p'_i d\theta_i}{(2\pi)^3\,2p_i^{\prime 0}} 
 =   \frac{\langle 12\rangle[21]}{2(2\pi)^3}\,da_{i1}da^*_{i1}da_{i2}da^*_{i2}.
\ee
Putting everything back together, the original integral can be written as
\bea
	I_{n\rightarrow2}=\frac{\left(\langle 12\rangle [21]\right)^{n-2}}{2^{n-2}n!(2\pi)^{3n-4}}\int\prod_{i=1}^{n}[ d^2a_{i1}d^2a_{i2}\,\, &   \delta\left(f_i(\a_i)\right)  f_i^\prime(\a_i) ]   \,   \delta^4\left(A^\dagger A -\mathds{1}_2\right)   \nonumber \\
	 &  \times \matrixel{12}{{\cal M}_{n\rightarrow2}}{1'\dots n'}\matrixel{1'\dots n'}{{\cal O}}{0}.\qquad\label{Eq:nphasesp}
\eea

\subsection{The $2\rightarrow 2$ case}

This is the simplest case. We would like to parametrize the matrix $A$ so that the delta function constraints take a particularly simple form. This can be achieved with a polar decomposition
\begin{equation}
	A = U P \, , \label{pdec}
\end{equation}
where $U$ is a unitary $2\times2$ matrix and $P$ is a positive-semidefinite hermitian matrix.
Without loss of generality we parametrise $P$ and $U$ as
\be
P=\begin{pmatrix}
	r & u+iv \\ u-iv & t
\end{pmatrix} \ , \quad 
U = \begin{pmatrix}
	e^{i\rho}\,\cos\theta & - e^{i\rho}\,\sin\theta\, e^{i\phi} \\
	e^{i\eta}\,\sin \theta\, e^{-i\phi} & e^{i\eta}\,\cos\theta
\end{pmatrix}, \label{paramap}
\ee 
where $r,\,u,\,v,\,t$ are real valued and  the angles have the ranges $0\le\rho,\,\eta,\,\phi<2\pi$ and $0\le\theta<\pi/2$.
The  delta function constraint  acts purely on the hermitian  matrix \reef{pdec}
\be
	\delta^4\left(A^\dagger A - \mathds{1}_2\right)   = \frac{1}{32}\delta(r-1)\delta(u)\delta(v)\delta(t-1) \,  .
\ee
Next, using \reef{paramap} and fixing  $P=\mathds{1}_2$, the delta function constraints that eliminate gauge redundancy take a simpler form 
 $
	\theta_1   = 2\rho +\phi \ , \quad 	\theta_2  = 2\eta - \phi 
 $. 
We can choose any topologically equivalent functions for the $f_i(\a_i)$, but it is convenient to pick the gauge where $f_1(\a_1)=\frac{1}{2}(\a_1-\phi)$ and $f_2(\a_2)=\frac{1}{2}(\a_2+\phi)$. The gauge fixing delta functions become
 $
\delta(f_1(\a_1))f^\prime_1(\a_1)\delta(f_2(\a_2))f^\prime_2(\a_2) = \frac{1}{4}\delta(\rho)\delta(\eta)
 $.
All in all, in terms of the new parameters, the integral measure becomes
\be
d^2a_{11}\,d^2a_{12}\,d^2a_{21}\,d^2a_{22} = 32 \sin 2\theta\,dr\,du\,dv\,dt\,d\rho\,d\eta\,d\theta\,d\phi,
\ee
once we have set $P=\mathds{1}_2$.
Putting everything together, the $2\rightarrow 2$ integral becomes
\be
	I_{2\rightarrow 2} =\frac{1}{16\pi}\int_{0}^{2\pi}\frac{d\phi}{2\pi}\int_{0}^{\pi/2}2\sin\theta\cos\theta d\theta\matrixel{12}{{\cal M}_{2\rightarrow 2}}{1'2'}\matrixel{1'2'}{{\cal O}}{0} \, . 
\ee

\subsection{The $3\rightarrow 2$ case}

Here we consider the phase space integral when 3 particles are exchanged across the cut. The calculation of the integral measure and transformation law for the spinors representing the exchanged particles proceeds in a similar way to the 2 particle case; the delta function constraints take a simpler form when the complex $3\times 2$ matrix $A$ is first expressed using a singular value decomposition, analogous to a polar decomposition for matrices that are not square
\be
A = U \Sigma V^\dagger \, . 
\ee
$U$ is a unitary $3\times 3$ matrix, $\Sigma$ is a $3\times 2$ rectangular diagonal matrix with real, non--negative entries, and $V$ is a unitary $2\times 2$ matrix. 
This decomposition is not unique however. It can be seen that applying the following transformation leaves $A$ unchanged:
\be
U\rightarrow U \;\text{diag}\{e^{i\phi_a},e^{i\phi_b},e^{i\phi_c}\},\qquad V\rightarrow V\; \text{diag}\{e^{i\phi_a},e^{i\phi_b}\} \, . 
\ee
To eliminate this redundancy, we choose to fix two of the undetermined phases in $V$ and one in $U$. With these restrictions, $U$ can be parametrized as
\be
U=\text{diag}\{e^{i\phi_2},e^{i\phi_3},e^{i\phi_4}\}
\begin{pmatrix}
	c_1 & -s_1c_3 & -s_1 s_3\\
	s_1c_2 & c_1c_2c_3 - s_2s_3 e^{i\delta} & c_1c_2s_3 + s_2c_3 e^{i\delta}\\
	s_1s_2 & c_1s_2c_3 + c_2s_3 e^{i\delta} & c_1s_2s_3 - c_2c_3 e^{i\delta}
\end{pmatrix}\text{diag}\{1,e^{i\phi},1\} \, , 
\ee
where $c_i\equiv\cos\theta_i$ and $s_i\equiv\sin\theta_i$. Note that alternative parametrizations of an arbitrary $3\times 3$ unitary matrix can be found by taking any parametrization of the CKM matrix and multiplying on either side by diagonal matrices of phases. We parametrize $\Sigma$  and $V$ (unitary matrix minus two redundant phases) as
\be \Sigma = 
\begin{pmatrix}
	r & 0 \\
	0 & t \\
	0 & 0
\end{pmatrix} \ , \quad  
V = \begin{pmatrix}
	c_4 & -s_4 e^{i\phi_5}\\
	s_4 e^{-i\phi_5} & c_4
\end{pmatrix} \, . 
\ee
In each case, the allowed ranges for the parameters is $0\le\theta_i<\pi/2$, $0\le r,t<\infty$ and $0\le \phi_i<2\pi$.
The delta function constraints depend only upon the parameters in $\Sigma$ and $V$:
\be
\delta\left(A^\dagger A-\mathds{1}_2\right) = \frac{1}{4 r t (r^2-t^2)^2 \sin2\theta_4}\delta(r-1)\delta(t-1)\delta(\theta_4)\delta(\phi_5) \, . 
\ee
The denominator is zero at the point in parameter space selected by the delta functions. This is just a harmless coordinate singularity though, which gets removed when combined with the integral measure in these new variables. This measure is
\be
\prod_{i=1}^{3}\left[d^2a_{i1}d^2a_{i2}\right] = 8r^3t^3(r^2-t^2)^2\cos\theta_1\sin^3\theta_1\sin2\theta_2\sin2\theta_3\sin2\theta_4\prod_{j=1}^{4}d\theta_j\prod_{k=1}^{5}d\phi_k\,dr\,dt\,d\delta  \, . 
\ee
Similarly to the $2\rightarrow2$ case, the gauge fixing delta functions become
\be
\delta(f_1(\a_1))f^\prime_1(\a_1)\delta(f_2(\a_2))f^\prime_2(\a_2)\delta(f_3(\a_3))f^\prime_3(\a_3) = \frac{1}{8}\delta(\phi_2)\delta(\phi_3)\delta(\phi_4) \, . 
\ee

Putting everything together into \reef{Eq:nphasesp}, yields the following result for the $I_{3\rightarrow2}$ integral
\be
I_{3\rightarrow 2}= \frac{\langle 12\rangle[21]}{4^4\pi^3 3!}\int d\Omega_3 \matrixel{12}{{\cal M}_{3\rightarrow 2}}{1'2'3'}\matrixel{1'2'3'}{{\cal O}}{0} \, , 
\ee
with the measure given by
\be
d \Omega_3 = 4\cos\theta_1\sin^3\theta_1d\theta_1 2\cos\theta_2\sin\theta_2d\theta_22\cos\theta_3\sin\theta_3 d\theta_3\,\frac{d\delta}{2\pi}\,\frac{d\phi}{2\pi} \, . 
\ee
After the all of the constraints have been applied, the transformation law for the spinors is
\be
\begin{pmatrix}
	\lambda^\prime_1 \\
	\lambda^\prime_2 \\
	\lambda^\prime_3 
\end{pmatrix} = 
\begin{pmatrix}
	c_1 & -e^{i\phi}c_3s_1 \\
	c_2s_1 & e^{i\phi}\left(c_1c_2c_3-e^{i\delta}s_2s_3\right) \\
	s_1s_2 & e^{i\phi}\left(c_1c_3s_2+e^{i\delta}c_2s_3\right)
\end{pmatrix}
\begin{pmatrix}
	\lambda_1\\
	\lambda_2
\end{pmatrix}.
\ee

%%%%%%%%%%%%%%%%%%%%%%%
%%%%%%%%                           %%%%%%%
%%%%%%%%   Amplitudes    %%%%%%%%
%%%%%%%%                           %%%%%%%
%%%%%%%%%%%%%%%%%%%%%%%%

\section{Amplitudes}

We computed some high point amplitudes of the text using the BCFW technique~\cite{Britto:2005fq}.~\footnote{Note that some of the SM amplitudes have a pole at infinity in the BCFW complex plane. In those cases we used standard Feynman diagram calculation.  }
In this appendix we illustrate how to compute the five point amplitude in \reef{ampffsgg}. 
An advantage of using BCFW w.r.t. Feynman diagrams  is that it automatically produces nice compact expressions  that simplify  our phase-space integrations. 

To apply BCFW recursion relations we need the   lower order seeds, namely the three point amplitudes. 
The three point amplitude between two fermions and a gluon is~\footnote{The normalisation is such that   the covariant derivative is $D_\mu = \partial_\mu + i T^A_{ij} G^A_\mu$.}
\be
M(1^-_i 2^+_j 3^-_A)\,=\, -g\sqrt{2}\,T^A_{ij}\,\frac{\ab{13}^2}{\ab{12}}
\quad \text{and} \quad 
M(1^-_i 2^+_j 3^+_A)\,=\, -g\sqrt{2}\,T^A_{ij}\,\frac{[13]^2}{[12]} \, , 
\ee
and for a Yukawa interaction $\mathcal{L}\supset - y H \bar{q}_L q_R$,
\be
M(1^- 2^- 3\,)\,=\, -y \ab{12}
\quad \text{and} \quad 
M(1^+ 2^+ 3\,)\,=\, -y[12] \, . 
\ee
In this case we do not  need  the three point amplitude between two scalars and  a gauge boson because the Higgs does not carry color.

To compute the five point amplitude \reef{ampffsgg} we need the   4-point amplitude  between two same-helicity fermions, a scalar and a gluon. It is easy to do it with the usual techniques, but we will do it using the BCFW recursion relation. See  \cite{Elvang:2015rqa} for a review.

BCFW proceeds by  shifting the spinors by a complex parameter $z$ so that the shifted momenta $\hat{p}$ is on-shell and total momentum is preserved, $\hat{p}^2=0$ and $\sum_ip^\mu_i=0$. 
 For instance, in a $n$-particle amplitude a $[1,2\rangle$-shift consists on shifting the spinors of $p_1$ and $p_2$ as
$| \hat{1} ] = | 1 ] + z |2]\,, 
| \hat{2} ] = | 2 ]\,,
| \hat{1} \rangle = | 1 \rangle\,,
| \hat{2} \rangle = | 2 \rangle - z | 1 \rangle$, and one can check that this preserves momentum conservation of on-shell momenta. For instance,  applying this shift on the $M(1^-_i 2^\pm_A 3^-_j 4\,)$   amplitude,   BCFW  splits it into two on-shell 3-point amplitudes
\be
 \begin{minipage}[h][1.1cm][t]{0.21\linewidth} \vspace{-.12cm} \dBCFWun \end{minipage}
 = 
  M_L(\hat{1}^-_i \hat{p}^-_k 4\,)\frac{1}{\ab{14}[14]} M_R(-\hat{p}^-_k 3^-_j \hat{2}^\pm_A)  \, . 
\ee
There are two different cases depending on the helicity of the gluon. 
If the   gluon has negative helicity, the numerator of the right amplitude is $\ab{3\hat{2}}^2$. However, this is evaluated at a $\hat{p}^2=0=\ab{\hat{2}3}[\hat{2}3]=\ab{\hat{2}3}[23]$. This means that $z$ is such that $\ab{\hat{2}3}=0$, and therefore  we recover that the all-minus amplitude vanishes $M(1^-_i 2^-_A 3^-_j 4\,)=0$. Instead, if the gluon has positive helicity  we get, 
\be
 M (1^-_i 2^+_A 3^-_j 4\,) = yg \sqrt{2} T^A_{ij}\, \ab{\hat{1}\hat{p}}\frac{1}{\ab{14}[14]}\frac{[\hat{p}\hat{2}]^2}{[\hat{p}3]} \, , 
\ee
where we used $\hat{p}=-\hat{p}_1-p_4$, $\ab{1\hat{p}}[\hat{p}2]=-\ab{14}[42]$. A basic  trick to get rid of $\hat{p}$ is to multiply the amplitude by $\ab{2\hat{p}}/\ab{2\hat{p}}$, and then write $\ab{2\hat{p}}[\hat{p}2]=\ab{23}[32]$ and $\ab{2\hat{p}}[\hat{p}3]=\ab{2\hat{2}}[23]$.  As we showed, this last factor has to be evaluated at $\ab{\hat{2}3}=0=\ab{23}-z\ab{13}$ so that $z=-\ab{23}/\ab{13}$. At this value, $\ab{2\hat{2}}=-z\ab{12}=\ab{12}\ab{23}/\ab{13}$. Then putting everything  together, we  get
\be
M(1^-_i 2^+_A 3^-_j 4\,) \,=\, - yg \sqrt{2} T^A_{ij}\, \frac{[24]^2}{[14][34]} 
\,=\, yg \sqrt{2} T^A_{ij}\, \frac{\ab{13}^2}{\ab{12}\ab{23}} \, .  \label{4ptrec}
\ee

Next we use the 4-point amplitude in  \reef{4ptrec} to construct   the 5-point amplitude $M(1^-_i 2^+_A 3^+_B 4^-_j 5\,)$. We do a  $[2,3\rangle$-shift. There are two contributions,
\be
 \begin{minipage}[h][1.1cm][t]{0.22\linewidth} \vspace{-.12cm} \dBCFWdos \end{minipage}+
  \begin{minipage}[h][1.1cm][t]{0.22\linewidth} \vspace{-.12cm} \dBCFWtres \end{minipage}
 \ee
 the two diagrams are given by
 \be
  M_L(1^-_i \hat{p}^+_k \hat{2}^+_A)\frac{1}{p_{12}^2}M_R(-\hat{p}^-_k \hat{3}^+_B 4^-_j 5\,)
\,+\,
M_L(1^-_i \hat{2}^+_A \hat{p}^-_k 5\,)\frac{1}{p_{34}^2} M_R(-\hat{p}^+_k \hat{3}^+_B 4^-_j ) \, . 
\ee
where we have  fixed a particular color ordering, namely   $(T^AT^B)_{ij}$. Exchanging $2\leftrightarrow 3$ in the sub-amplitudes  generates the other color ordering $(T^BT^A)_{ij}$. The first diagram vanishes because   it is evaluated at $[1\hat{2}]=0$, and $|\hat{p}\rangle[\hat{p}\hat{2}] = (p_1+\hat{p}_2)|\hat{2}\rangle = -|1\rangle[1\hat{2}]=0$. Hence  $[\hat{p}\hat{2}]=0$ as well, and therefore the left amplitude vanishes. The second diagram gives
\be
M(1^-_i 2^+_A 3^+_B 4^-_j 5\,) \,=\, -2 y g^2 (T^AT^B)_{ij}\,\frac{\ab{1\hat{p}}^2}{\ab{1\hat{2}}\ab{\hat{2}\hat{p}}} \frac{1}{\ab{34}[34]}\frac{[\hat{p}\hat{3}]^2}{[\hat{p}4]} \, . 
\ee
The same argument does not work for this diagram, since the right amplitude is proportional to $[\hat{p}\hat{3}]$ but instead it is evaluated at $\ab{\hat{3}4}=0$. Now, it is straightforward to simplify the expression using  $\ab{1\hat{p}}[\hat{p}\hat{3}]=\ab{14}[43]$ and $\ab{\hat{2}\hat{p}}[\hat{p}4]=\ab{2\hat{3}}[34]$; also, note that $\ab{2\hat{3}}=\ab{23}$. Therefore the 5-point amplitude is  
\be
M(1^-_i 2^+_A 3^+_B 4^-_j 5\,) \,=\, -2 y g^2 (T^AT^B)_{ij}\,\frac{\ab{14}^2}{\ab{12}\ab{23}\ab{34}}\,
-2 y g^2 (T^BT^A)_{ij}\,\frac{\ab{14}^2}{\ab{13}\ab{32}\ab{24}} \,  .
\ee
The amplitude can be crosschecked versus the $N=4$ SYM Parke-Taylor, up to color factors.

%%%%%%%%%%%%%%%%%%%%%%%%%%%
%%%%%%%%                           %%%%%%%%%%%
%%%%%%%%   Bibliography    %%%%%%%%%%%
%%%%%%%%                           %%%%%%%%%%%
%%%%%%%%%%%%%%%%%%%%%%%%%%%

\small

%\bibliography{biblio}
\bibliographystyle{utphys}

\end{document}